# High-Resolution Observations of Bright Boulders on Asteroid Ryugu: 2. Spectral Properties

**Chiho Sugimoto[1], Eri Tatsumi[1,2], Yuichiro Cho[1], Tomokatsu Morota[1], Rie Honda[3], Shingo Kameda[4], Yosuhiro Yokota[3,5], Koki Yumoto[1], Minami Aoki[1], Daniella N. DellaGiustina[6,7], Tatsuhiro Michikami[8], Takahiro Hiroi[9], Deborah L. Domingue[10], Patrick Michel[11], Stefan Schröder[12], Tomoki Nakamura[13], Manabu Yamada[14], Naoya Sakatani[4], Toru Kouyama[15], Chikatoshi Honda[16], Masahiko Hayakawa[5], Moe Matsuoka[5], Hidehiko Suzuki[17], Kazuo Yoshioka[1], Kazunori Ogawa[18], Hirotaka Sawada[5], Masahiko Arakawa[15], Takanao Saiki[5], Hiroshi Imamura[5], Yasuhiko Takagi[19], Hajime Yano[5], Kei Shirai[15], Chisato Okamoto[15†], Yuichi Tsuda[5], Satoru Nakazawa[5], Yuichi Iijima[5†], and Seiji Sugita[1,14,*]**

[1] The University of Tokyo, Tokyo 113-0033, Japan.

[2] Instituto de Astrofísica de Canarias (IAC), 38205 La Laguna, Tenerife, Spain.

[3] Kochi University, Kochi 780-8520, Japan.

[4] Rikkyo University, Tokyo 171-8501, Japan.

[5] Institute of Space and Astronautical Science (ISAS), Japan Aerospace Exploration Agency (JAXA), Sagamihara 252-5210, Japan.

[6] Lunar and Planetary Laboratory, University of Arizona, Tucson, AZ 85721, USA.

[7] Department of Geosciences, University of Arizona, Tucson, AZ 85721, USA.

[8] Faculty of Engineering, Kindai University, Higashi-Hiroshima 739-2116, Japan.

[9] Department of Earth, Environmental and Planetary Sciences, Brown University, Providence, RI 02912, USA.

[10] Planetary Science Institute, Tucson, AZ 85719, USA.

[11] Université Côte d'Azur, Observatoire de la Côte d'Azur, Centre National de le Recherche Scientifique, Laboratoire Lagrange, 06304 Nice, France.

[12] German Aerospace Center (DLR), Institute of Planetary Research, 12489 Berlin, Germany.

[13] Department of Earth Science, Tohoku University, Sendai 980-8578, Japan.

[14] Planetary Exploration Research Center, Chiba Institute of Technology, Narashino 275-0016, Japan.

[15] National Institute of Advanced Industrial Science and Technology, Tokyo 135-0064 Japan.

[16]University of Aizu, Aizu-Wakamatsu 965-8580, Japan.

[17] Meiji University, Kawasaki 214-8571, Japan.

[18] Kobe University, Kobe 657-8501, Japan.

[19]Aichi Toho University, Nagoya 465-8515, Japan

[†] Deceased.

[*] Corresponding author.

E-mail address: sugita@eps.s.u-tokyo.ac.jp

*Abstract:* Many small boulders with reflectance values higher than 1.5 times the average reflectance have been found on the near-Earth asteroid 162173 Ryugu. Based on their visible wavelength spectral differences, Tatsumi et al. (2021, Nature Astronomy, 5, doi:https://doi.org/10.1038/s41550-020-1179-z) defined two bright boulder classes: C-type and S-type. These two classifications of bright boulders have different size distributions and spectral trends. In this study, we measured the spectra of 79 bright boulders and investigated their detailed spectral properties. Analyses obtained a number of important results. First, S-type bright boulders on Ryugu have spectra that are similar to those found for two different ordinary chondrites with different initial spectra that have been experimentally space weathered the same way. This suggests that there may be two populations of S-type bright boulders on Ryugu, perhaps originating from two different impactors that hit Ryugu's parent body. Second, the model space-weathering ages of meter-size S-type bright boulders, based on spectral change rates derived in previous experimentally irradiated ordinary chondrites, are $10^5$–$10^6$ years, which is consistent with the crater retention age ($<10^6$ years) of the ~1-m deep surface layer on Ryugu. This agreement strongly suggests that Ryugu's surface is extremely young, implying that the samples acquired from Ryugu's surface should be fresh. Third, the lack of a serpentine absorption in the S-type clast embedded in one of the large brecciated boulders indicates that fragmentation and cementation that created the breccias occurred after the termination of aqueous alteration. Fourth, C-type bright boulders exhibit a continuous spectral trend similar to the heating track of low-albedo carbonaceous chondrites, such as CM and CI. Other processes, such as space weathering and grain size effects, cannot primarily account for their spectral variation. Furthermore, the distribution of the spectra of general dark boulders, which constitute >99.9% of Ryugu's volume, is located along the trend line in slope/UV-index diagram that is occupied by C-type bright boulders. These results indicate that thermal metamorphism might be the dominant cause for the spectral variety among the C-type bright boulders on Ryugu and that general boulders on Ryugu may have experienced thermal metamorphism under a much narrower range of conditions than the C-type bright boulders. This supports the hypothesis that Ryugu's parent body experienced uniform heating due to radiogenic energy rather than impact heating.

Highlights:

We conducted spectroscopic analysis of newly found >70 bright boulders. (71)

S-type bright boulders on Ryugu follow two different space-weathering tracks. (79)

Ryugu's parent body may have been hit by more than one large projectile. (75)

Largest S-type clast embedded in large breccia indicated no serpentine absorption. (85)

Spectral trend of C-type bright boulders resembles CM/CI heating tracks. (79)

## 1. Introduction

Initial global observations of the asteroid Ryugu by the Japanese spacecraft Hayabusa2 by Japan Aerospace Exploration Agency (JAXA) show that Ryugu has many spots much brighter than average Ryugu surfaces (Sugita et al., 2019). They were subsequently found to be boulders and then named as "bright boulders" (Tatsumi et al., 2021).Tatsumi et al. (2021) further found that these bright boulders can be divided into two spectral types (C-type and S-type) based on principal-component analysis of their visible 5-band spectra. Note that we refer to C-/X-type bright boulders in Tatsumi et al. (2021) as C-type bright boulders in this paper, because of their similarity to carbonaceous chondrites as discussed in Paper 1. To determine the detailed properties of these bright boulders on Ryugu, Sugimoto et al. (2021) conducted a series of image analyses with high-resolution data obtained during special spacecraft operations associated with touchdown rehearsals and the search for the artificially created crater made during the Small Crater Impact (SCI) experiment. These operations took the spacecraft down to 80 m altitude during the rehearsal operations and down to 1.7 km of altitude in the search for the SCI crater. We refer to the companion paper by Sugimoto et al. (2021) as Paper 1 hereinafter. Paper 1 found more than a thousand bright boulders and measured their size distribution. Using high-resolution images (i.e., down to 3.8 mm/pix), Paper 1 investigated the detailed morphologies of bright boulders and found that many of them are polymict breccias. However, the paper did not address the detailed spectral properties of these newly found bright boulders on Ryugu, which we address in this study.

High-resolution observations allow us to investigate the spectral variations within bright boulders. Color variations among bright boulders within the same spectral type helps constrain their origins. For example, the relatively large variety of visible spectra among S-type bright boulders on Ryugu found by Tatsumi et al. (2021) may reflect the presence of fragments from multiple projectiles. However, this spectral variety might also be due to the different degrees of spectral modification of the same original material. Thus, in this study, we examine this spectral variety among S-type bright boulders on Ryugu and constrain its possible origin.

The color variation among Ryugu's C-type bright boulders has not been investigated by prior studies thus far. Although the number of C-type bright boulders analyzed by Tatsumi et al. (2021) are greater than that of S-types, their great spectral variety renders it difficult to determine whether their variation is one continuous distribution or multiple populations. Thus, many other C-type bright boulders must be analyzed. Furthermore, although their distribution ranges are consistent with the spectral variation of some carbonaceous chondrites that have experienced different degrees of dehydration, no other likely processes, such as space weathering or grain size effects, have been examined extensively thus far. The spectral variety among the C-type bright boulders also helps constrain the processes that led to their formation. If bright boulders are fragments of the parent body, their spectral variety should be explained by some spectral variation processes, including thermal metamorphism, aqueous alteration, and space weathering, on materials constituting Ryugu. In contrast, if their spectral variation is too large to be derived from the same parent body, the exogenous origin of the C-type boulders would be indicated in a similar manner as S-type bright boulders on Ryugu and pyroxene-rich boulders on Bennu (DellaGiustina et al. 2021 and Tatsumi et al. 2021). In this study, we investigate the variation in the spectral properties of C-type bright boulders and examine whether they can be derived from the same material that constitutes the bulk mass of Ryugu.

The remainder of this paper is organized as follows. We discuss the data analysis in Section 2, spectral analysis results in Section 3, spectral modification mechanisms on Ryugu's materials and geological implications for the evolution of Ryugu and its parent body in Section 4, and conclusions in Section 5.

## 2. Method

Accurate spectral analysis of bright boulders requires the accurate separation of photons coming from a bright boulder and those surrounding the substrate boulder/background regolith bed. Because the side lengths of about a half of bright boulders analyzed in this study are ≤3 pixels of the charge-coupled device's (CCD) on the Optical Navigation Camera telescope (ONC-T) onboard Hayabusa2, an accurate size estimation of these bright boulders requires additional analysis (i.e., profile-based size estimation) for separation between bright boulders and background. In this section, we first discuss the multi-band image data reduction in Section 2.1, the profile-based size estimation methodology in Section 2.2, the extraction of spectra of small bright boulders in Section 2.3, and the photometric correction in Section 2.4.

### 2.1 Datasets and processing

We used images in seven filters (ul: 0.40 $\mu$m, b: 0.48 $\mu$m, v: 0.55 $\mu$m, Na: 0.59 $\mu$m, w: 0.70 $\mu$m, x: 0.86 $\mu$m, and p: 0.95 $\mu$m) and four band images (ul, b, v, and x bands) captured by ONC-T onboard Hayabusa2 (Kameda et al., 2017; Sugita et al., 2019). The data we analyze in this study are the same as in Paper 1 except that Paper 1 analyzed only v-band images. More specifically, we analyzed the same 12 sets of four-band images from artificial-crater-search observations (CRA1 and CRA2) at ~1.7 km of altitude as Paper 1. More details, such as observation time and footprint on Ryugu, of these images are given in Paper 1. We also use the same high-resolution images taken at low altitudes as Paper 1 and medium-resolution images taken at 2.7 km after MASCOT release observations studied by Tatsumi et al. (2021). Data processing, such as the removal of stray light, bias, and read-out smear, flat fielding, and I/F conversion, are also the same as in Paper 1. Among bright boulders newly found by the automatic detection method by Paper 1, only 79 of them have 4-filter coverage by CRA1 and CRA2 observations. These bright boulders found within CRA1 and CRA2 images have been given ID numbers from C1 through C79, separated from the M series found in MASCOT hovering images by Tatsumi et al. (2021) because these two sets of images were obtained under very different observational conditions, such as resolution and viewing geometry. The details, such as location (i.e., latitude and longitude), size, and spectral type, of the new C-series bright boulders are given in Table 1 in Paper 1. We analyzed the spectra of C1 – C79 boulders using the method described in section 2.2, 2.3, and 2.4.

## 2.2 Profile-based size estimation of small bright boulders

The sizes of small (< 3 pixels in side length) bright boulders, which are too small for direct size measurement based on images, were estimated using the reflectance profile of each bright boulder in v-band I/F images. For small boulders <3 pixels across, because the Point Spread Function (PSF) with the Full Width at Half Maximum (FWHM) is ~1.7 pixels (Kameda et al., 2017), the rims of these small bright boulders are blurred. Furthermore, although the actual size of the target is smaller than 1 pixel, it can overlap with up to four CCD pixels if it is located near the corner of a CCD pixel. Thus, we created a series of simulated images in which an artificial bright boulder is blurred using a gaussian filter and compared the relationship between the size of the artificial bright boulder and its apparent FWHM with the small bright boulder images. The use of gaussian is appropriate because PSF analysis by Kameda et al. (2017) is based on gaussian fitting.

We created images of squares and circles with a side length or a diameter of 0.1–6.9 pixels in increments of 0.1 pixels whose upper-left corner is located at the upper-left corner of a CCD pixel. An example of a square with a side length of 1.5 pixels is shown in Fig. 1A. Artificial bright boulders were given a brightness value of 1 and a background of 0. Squares were shifted by -0.5 to +0.5 pixel in steps of 0.1 pixels in the x- or/and y-direction (Fig. 1B) and blurred with a gaussian filter with $\sigma = 1.78/2\sqrt{2\ln 2}$, where $2\sqrt{2\ln 2}$ is the ratio of FWHM to $\sigma$ of the Gaussian distribution (Fig. 1C). Note that one pixel was divided into 100 grids in this study; each cell in the grey mesh shown in Figs. 1A-C was divided into $100 \times 100$ grid elements. The image was then binned to each pixel (Fig. 1D). The brightness profile across the brightest point of the binned image was acquired and normalized at the peak (Fig. 2). Using linear interpolation, we determined the FWHM. The same experiment was conducted for circles with diameters of 0.1–6.9 pixels over size increments of 0.1 pixels.

Figure 3A shows the relationship between the side length or diameter of an artificial bright boulder and its FWHM. Square and circle artificial bright boulders with sizes (i.e., side length or diameter) smaller than ~1 pixel show almost the same distribution. In contrast, square artificial bright boulders show wider FWHM than circular ones with the same size in larger size ranges. Although there are gaps in the FWHM distribution between squares and circles in some size ranges, an intermediate shape between circles and squares is expected to fill these gaps. For each size, the range of FWHM that an artificial bright boulder can take is ~ 0.4 pixels. As for larger sizes (>3 pixels), this gap width is attributed to the difference in shape. In contrast, as for smaller sizes (<3 pixels), this gap is caused by the sub-pixel shifting of the target. Furthermore, in this smaller size range, FWHMs of artificial bright boulders are much wider than their sizes. This suggests that objects with sizes smaller than ~3 pixels are substantially affected by the PSF in the ONC-T images.

The black dashed lines in Fig. 3 show the spline fitting of upper and lower limits of the FWHM distribution. The lower limit line converges to FWHM of ~1.9 pixels, which is about the width of the PSF. The upper-limit line converges to an FWHM of ~2.3 pixels, which is about the minimum FWHM for an artificial bright boulder with a size of 2 pixels. This suggests that the widest FWHM occurs when the center of an artificial bright boulder is located at the boundary of a CCD pixel and the light intensity of an artificial bright boulder is evenly divided into two pixels. The sizes of observed bright boulders were estimated in this study by comparing these upper and lower limit lines with the FWHMs of observed bright boulders.

The FWHMs of observed bright boulders were measured in v-band I/F images in the following way. First, we obtained the reflectance profiles across the brightest point of the observed bright boulders. Reflectance profiles in both x- and y-directions were examined for each bright boulder. Figure 4 shows an example of bright boulder analyzed in this study. Figure 5A is the x-direction reflectance profile of the bright boulder shown in Fig. 4A. Second, we subtracted the base intensity (i.e., radiance of the background) from the reflectance profile and then divided it by the height $h$ of the peak. The radiance of the background was estimated from the surrounding area of each bright boulder, shown as the larger square area with a size of 441 pixels (21 pixels in side length), excluding the smaller square with a size of 49 pixels (7 pixels in side length) in Fig. 4B. Using these 392 pixels, we calculated the Median Absolute Deviation (MAD). MAD is calculated as the median of absolute deviations between data values and their median. For a normal distribution, MAD is equal to 1.48 times the standard deviation (STD) $\sigma$ and is a more robust estimation of the dispersion of data values, where outliers, such as shadows, are present in the dataset. Thus, we removed pixels whose reflectance is $2\sigma \sim 2.96\ \mathrm{MAD}$ darker than the median of the 392 pixels and then recalculated the median and MAD of the surrounding area. In Fig. 4B, the area used to calculate the median of the background is shown as a gray mask. The mask has nulls that are pixels removed from shadows. The FWHM of a normalized reflectance profile depends on the base intensity. Considering the error in the median value of the background reflectance (i.e., STD $\sim 1.48\,\mathrm{MAD}$), we measured the upper and lower limits of the FWHM of the normalized reflectance profile: the width between points where the normalized reflectance profile equals $0.5 - 1.48\,\mathrm{MAD}/h$ (minimum of FWHM) and $0.5 + 1.48\,\mathrm{MAD}/h$ (maximum of FWHM) (Fig. 5B). Comparing these maximum and minimum values of observed FWHM with the lower- and upper-limit lines of the artificial bright boulders, the maximum and minimum values of the bright boulder size were estimated (Fig. 3B). Note that the contribution of shot noise in the normalized reflectance profile was smaller than ~1 % and is negligible in this study. A similar assessment for larger shot noise in other situations is discussed in detail below in section 2.3.

When the minimum FWHM of the observed bright boulders is smaller than ~2.3 pixels, this estimation provides the minimum size of < 0.1 pixels. However, bright boulders must have finite sizes to be recognized as bright boulders. Thus, we estimated the minimum sizes of bright boulders to reproduce the brightest pixel. When the observed brightest pixel is $\alpha$ times brighter than the reflectance $r$ of the background and the material of the bright boulder is $\beta$ times brighter than the background, the following formula can be established:

$$\alpha r = r(1 - x) + \beta r x \,, \tag{2.2.1}$$

where $x$ is the area ratio of bright boulder to 1 pixel. Solving for $x$, this equation provides the dimension $\sqrt{x}$ of a bright boulder by:

$$\sqrt{x} = \sqrt{\frac{\alpha - 1}{\beta - 1}} \,. \tag{2.2.2}$$

If we assume a reflectance for a bright boulder, we can then estimate the minimum size of such a bright boulder. The majority of the surface of Ryugu is spectrally very uniform and extremely dark (~4.0 % of normal albedo; Tatsumi et al., 2020). It is noted that two types (C-type and S-type) of bright boulders are observed on Ryugu, and that S-type (ordinary-chondrite-like) bright boulders tend to be brighter than the other (Tatsumi et al., 2021). Because larger brightness ratio $\beta$ value gives smaller $\sqrt{x}$ as Eq. 2.2.2 shows, the lower limit of the bright boulder size is obtained when $\beta$ of an S-type bright boulder, which is greater than a C-type one, is used. Typical normal albedo of ordinary chondrites is ~50% (Piironen et al., 1998). Thus, in this study, to estimate the lower limit of the finite size of a bright boulder, we assumed that the material of bright boulders are ten times ($\beta$=10) brighter than the background (i.e., the Ryugu surface) or lower.

Combining the results of the size estimations based on the reflectance profiles and the constraint for the smallest size based on the brightness value of the brightest pixel, maximum ($x_{max}, y_{max}$) and minimum ($x_{min}, y_{min}$) values of x- and y-direction sizes were determined for the set of boulders in our study. The x- and y-direction sizes ($x_{ave}, y_{ave}$) were then calculated as the average of maximum and minimum values. Using the average, maximum, and minimum x- and y-direction sizes, the average, maximum, and minimum areas ($S$, $S_{max}$, $S_{min}$, respectively) of the bright boulders were calculated using:

$$S = x_{ave}\, y_{ave}, \tag{2.2.3}$$

$$S_{max} = \frac{\pi}{4}\, x_{max}\, y_{max}, \tag{2.2.4}$$

$$S_{min} = x_{min}\, y_{min}. \tag{2.2.5}$$

Note that the maximum sizes of the artificial bright boulders were found for circular boulders; thus, Eq. 2.2.4 for the maximum area include a factor of $\pi/4$. Upper and lower errors ($\Delta S_1$ and $\Delta S_2$, respectively) of the area are given by

$$\Delta S_1 = S_{max} - S, \tag{2.2.6}$$

$$\Delta S_2 = S - S_{min}. \tag{2.2.7}$$

This size estimate is used in spectral extraction method B as discussed below in section 2.3.

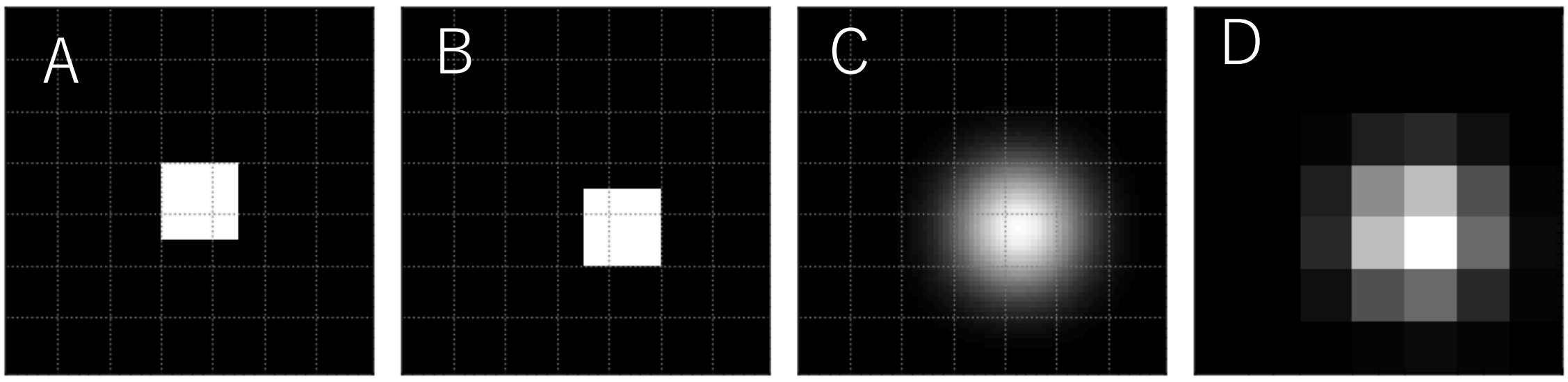

**Fig. 1** (A) Artificial bright boulders expressed as a square with a side length of 1.5 pix. Dotted gray lines indicate the boundary of CCD pixels. (B) An artificial bright boulder was shifted 0.5 pixel in both x- and y-directions from the position shown in (A). (C) Artificial bright boulder blurred by the Gaussian filter. (D) Binned image of (C).

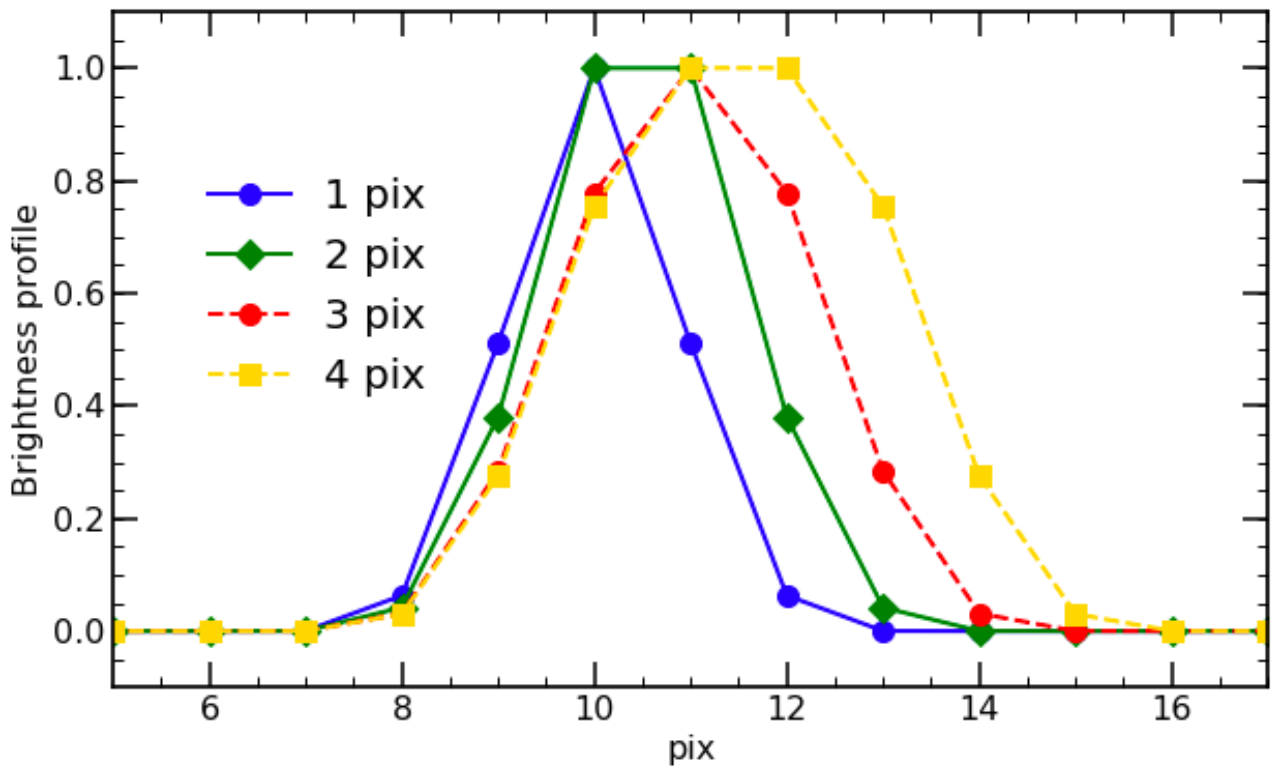

**Fig. 2** Brightness profiles of artificial bright boulders. The results for squares with a side length of 1, 2, 3 and 4 pixels are shown. Artificial bright boulders were blurred by a Gaussian filter and binned to individual CCD pixels. The brightness profiles were normalized to their brightest pixel.

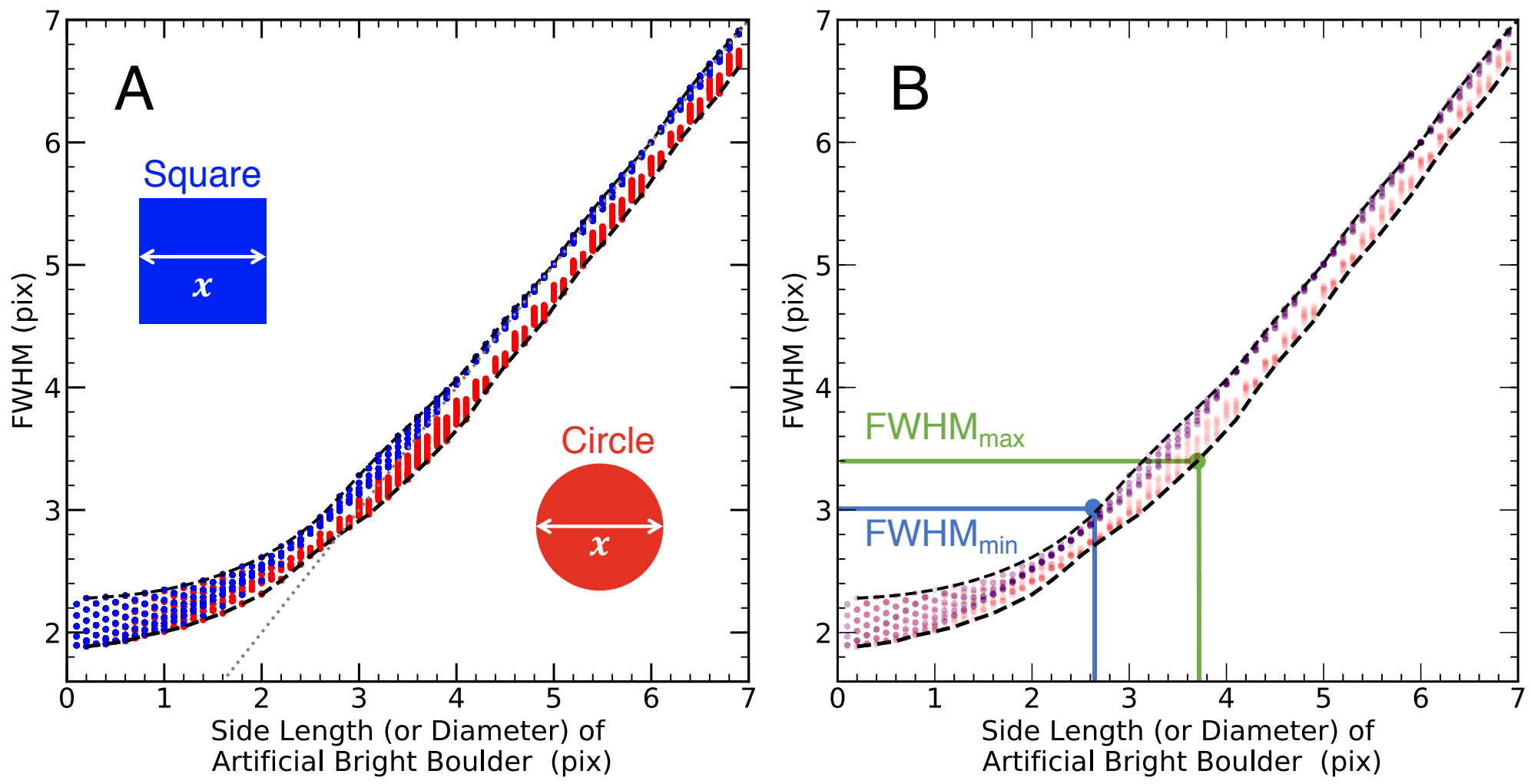

**Fig. 3** (A) Relationship between the sizes (i.e., side length or diameter) of artificial bright boulders and their FWHMs. (A) Blue and red plots indicate square and circle shaped simulated bright boulders, respectively. The gray dashed line is the line on which the FWHM is equal to the sizes. Note that many blue and dots are overlapping each other. (B) Black dashed lines represent the spline fittings of the upper and lower limits of the FWHMs. The spline fit lines were used to estimate the sizes of the observed bright boulders (modified from Tatsumi et al., 2021).

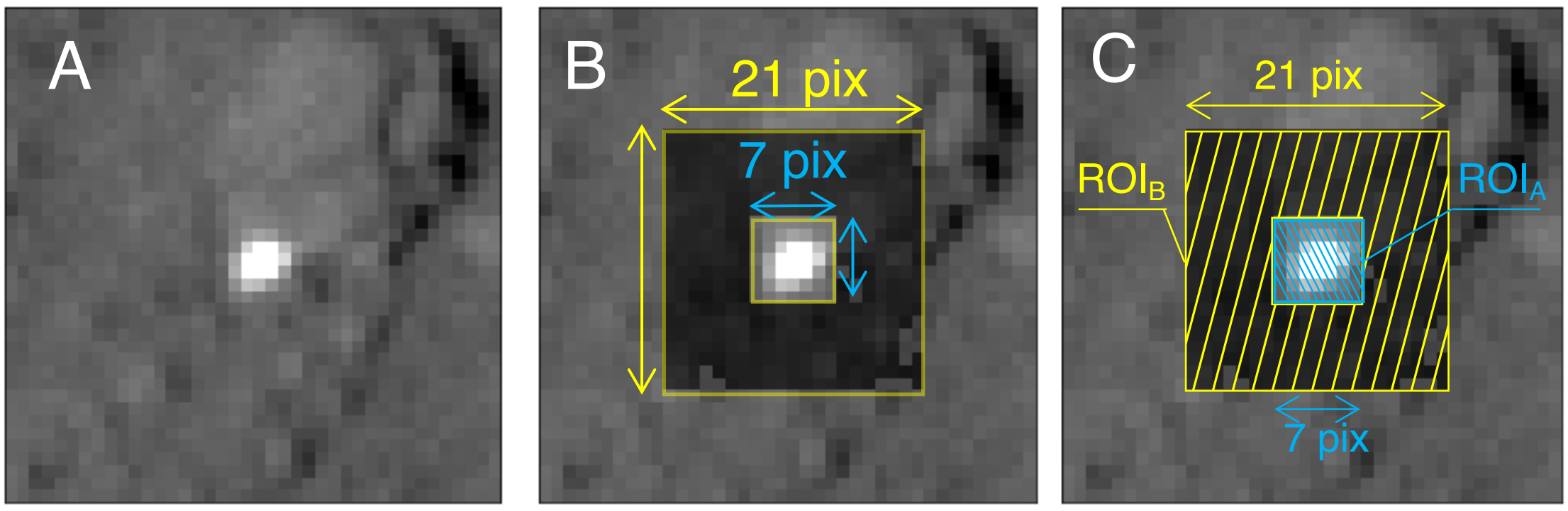

**Fig. 4** (A) v-band I/F image of M7 (C65) captured on October 3, 2019. (B) The gray mask indicates the area used to estimate the median reflectance of the background. The size of the outer larger square (indicated by the yellow arrows) is 21 pixels in side length, and that of the inner smaller square (indicated by the cyan arrows) is 7 pixels in side length. The gray mask is the larger square excluding the smaller square and pixels attributed to shadows. (C) Regions of interest (ROI) used to extract the radiance of bright boulders. The $ROI_A$ is the smaller cyan square area with a side length of 7 pixels. The typical area of $ROI_A$ is 49 pixels before shadow pixels were removed. A bright boulder is located at the center of this area. $ROI_B$ indicates the larger yellow square, excluding the cyan square. The typical area of $ROI_B$ is 392 pixels, before shadow pixels were removed. The average radiance of the background area adjacent to the bright boulder was derived from this area and subtracted from the total radiance of $ROI_A$.

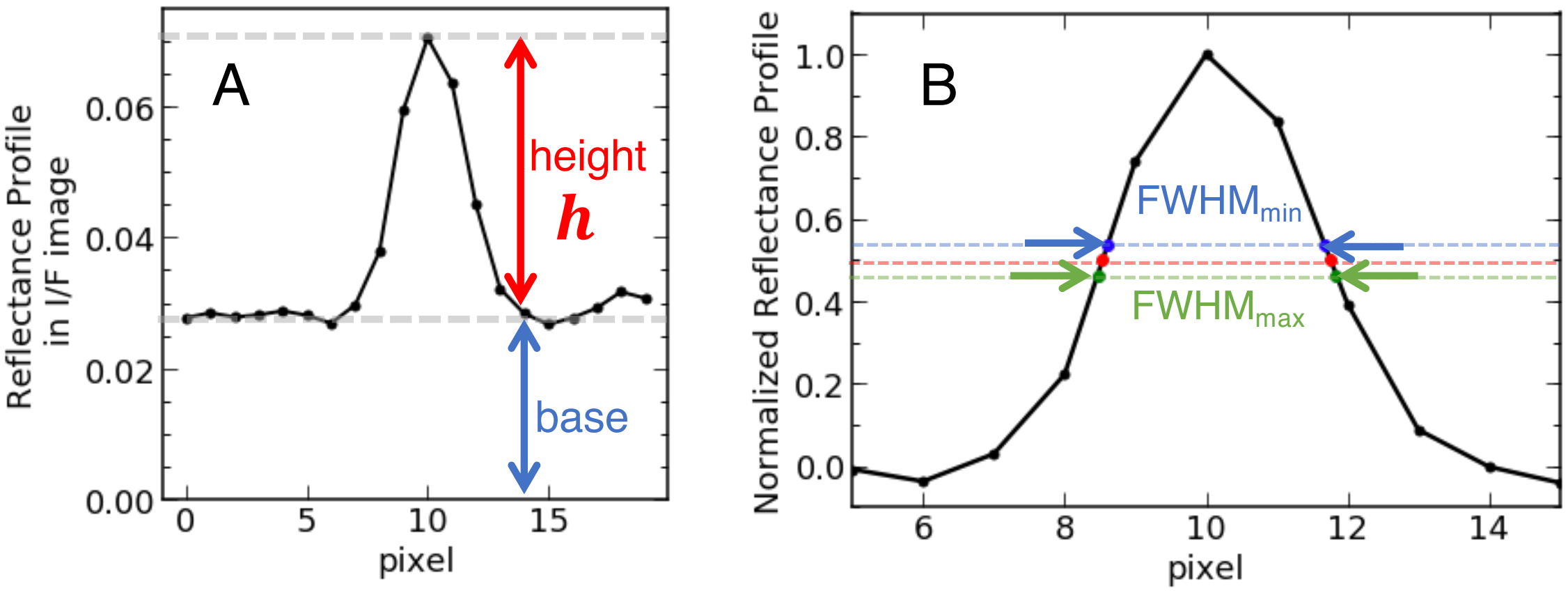

**Fig. 5** (A) The x-direction reflectance profile of the bright boulder shown in Fig. 4A. The base is the median reflectance of the surrounding area, excluding the shadows , as shown in Fig. 4B as the gray mask. The profile was derived by subtracting the base height and then dividing by the height of the peak of the bright boulder for normalization. (B) Normalized reflectance profile. In this normalized profile, the error in the base value (i.e., STD ~1.48 MAD) corresponds to $1.48\ \mathrm{MAD}/h$. Thus, the width of the points where the normalized profile equals $0.5 - 1.48\ \mathrm{MAD}/h$ and $0.5 + 1.48\ \mathrm{MAD}/h$ was measured as the maximum (green dashed line) and minimum (blue dashed line) of the FWHM.

## 2.3 Extraction of spectra

As objects with a side length smaller than ~2 pixels cannot be evaluated directly from the images due to blurring by the PSF, we evaluated the spectra in either one of the following two ways, depending on the pixel sizes of bright boulders. For bright boulders with areas > 9 pixel$^2$ (i.e., diameter > 3 pixels), we estimated their spectra by averaging the reflectance of the pixels within the area occupied by the boulder for each band (Method A in Table 1 in Paper 1). In the study by Tatsumi et al. (2021), the sizes of the integrated areas were adjusted to the rims of the bright boulders. However, the boundaries of some bright boulders, such as M2 in the Extended Figure 1 in Tatsumi et al. (2021), are unclear when they are interpreted as a bright subsection within another boulder, or a bright clast. Thus, to evaluate them more fairly, we assessed the spectra with an ROI of 9 pixel$^2$ (i.e., square area with side length of 3 pixels), where the brightest points are located at the center. The error in each filter band for each bright boulder was given by the standard error of I/F within the ROI. For bright boulders with areas ≦ 9 pixel$^2$ (corresponding to the diameter ≦ 3 pixels), we extracted the spectra of these bright boulders using the following method (Method B in Table 1 of Paper 1). First, using the same method in Section 2.2, we removed pixels occupied by shadows. Using a square area with a size of 21×21 pixel$^2$ (excluding the square area with a size of 7×7 pixel$^2$, Fig. 4C), we calculated the median and MAD. A bright boulder is located at the center of these squares in each band image. We removed pixels whose reflectance is darker than the median by $2\sigma$ (i.e., 2.96 MAD) in at least one band image.

When the areas adjacent to the bright boulder comprise many shadows, the above procedure sometimes cannot remove all the shadows. Obvious shadows, such as dark areas next to the boulders, are left out in the processed image. Then, we removed pixels whose reflectance is darker than the median by 1 $\sigma$ (i.e., 1.48 MAD), as opposed to 2 $\sigma$. These shadow criteria were applied for some bright boulders (e.g., C32, C35, C40, C42, C43, C49, C52, and C63) observed during CRA1 and CRA2. Due to their higher spatial resolution and larger phase angles compared with the observations during the hovering operation after the MASCOT deployment, CRA1 and CRA2 images tend to have a larger pixel area of shadows.

Subsequently, the radiance of a bright boulder was calculated by subtracting the contribution of the background (the adjacent area of bright boulders) from the total radiance of the area around the bright boulder ($ROI_A$ in Fig. 4C). Using the area of bright boulders estimated by the method described in section2.2 $S$, the radiance of the bright boulder $F$ is given by:

$$F = \frac{I_A - I_{bg}}{S} = \frac{I_A - i_B(L - S)}{S}\ , \tag{2.3.1}$$

where $I_A$ is the total radiance in $ROI_A$, $L$ is the area of $ROI_A$, and $I_{bg}$ is the estimated total radiance contaminated from the background. Background radiance $I_{bg}$ was estimated from the surrounding area of $ROI_A$ ($ROI_B$ in Fig.4C): $I_{bg} = i_B(L - S)$, where $i_B$ is the mean radiance of $ROI_B$. The uncertainty in the radiance of the bright boulders was evaluated using the error propagation in three terms: the error for the total radiance in $ROI_A$, the area, and the mean radiance of $ROI_B$. Area errors $\Delta S$ are expressed by Eqs. (2.2.6) and (2.2.7). The error $\Delta I_A$ for the total radiance in $ROI_A$ is given as the shot noise in the $ROI_A$. Using the gain factor for the ONC-T of 20.95 $e^-$/DN (Kameda et al., 2017), the shot noise is given by:

$$\Delta I_A = \sqrt{\frac{N}{20.95}}\frac{\pi}{S_n F_{solar} t}\left(\frac{d}{1\ AU}\right)^2, \tag{2.3.2}$$

where $N$ is total count values in the $ROI_A$, $S_n$ is the sensitivity of each band filter, $F_{solar}$ is solar irradiance, $t$ is exposure time, and $d$ is the distance between Ryugu and the Sun. The factor $\sqrt{N/20.95}$ is the digital count values of the shot noise when N is a 12-bit digital number. The following factor $\frac{\pi}{S_n F_{solar} t}\left(\frac{d}{1\ AU}\right)^2$ is used to convert it to a radiance factor. The error for the mean radiance of the $ROI_B$, $\Delta i_B$, is given as the standard error of the $ROI_B$. Error coefficients for each factor are given by:

$$\frac{\partial F}{\partial I_A} = \frac{1}{S}, \tag{2.3.3}$$

$$\frac{\partial F}{\partial i_B} = -\frac{L - S}{S}, \tag{2.3.4}$$

$$\frac{\partial F}{\partial S} = -\frac{I_A - i_B L}{S^2}\ . \tag{2.3.5}$$

Thus, the error, $\Delta F$, for the radiance of the bright boulders is obtained from the error propagation.

$$\Delta F = \sqrt{\left(\frac{\partial F}{\partial I_A}\Delta I_A\right)^2 + \left(\frac{\partial F}{\partial i_B}\Delta i_B\right)^2 + \left(\frac{\partial F}{\partial S}\Delta S\right)^2} \tag{2.3.6}$$

The normalized reflectance $f_i$ of the i-th band ($\mathrm{i = ul, b, Na, w, x, p}$) is given by

$$f_i = \frac{F_i}{F_v} = \frac{I_{Ai} - i_{Bi}(L - S)}{I_{Av} - i_{Bv}(L - S)}, \tag{2.3.7}$$

where $I_{Ai}$ and $i_{Bi}$ are the total radiance in the $\mathrm{ROI_A}$ and the mean radiance of the $\mathrm{ROI_B}$ in i-th band, respectively, and $I_{Av}$ and $i_{Bv}$ are the values in the v-band. Error coefficients for each factor are given by:

$$\frac{\partial f_i}{\partial S} = \frac{i_{Bi}\{I_{Av} - i_{Bv}(L - S)\} - i_{Bv}\{I_{Ai} - i_{Bi}(L - S)\}}{\{I_{Av} - i_{Bv}(L - S)\}^2} \tag{2.3.8a}$$

$$= \frac{-I_{Av} i_{Bv}\left(\frac{I_{Ai}}{I_{Av}} - \frac{i_{Bi}}{i_{Bv}}\right)}{\{I_{Av} - i_{Bv}(L - S)\}^2} \tag{2.3.8b}$$

$$\frac{\partial f_i}{\partial i_{Bi}} = \frac{(L - S)}{I_{Av} - i_{Bv}(L - S)} \tag{2.3.9}$$

$$\frac{\partial f_i}{\partial i_{Bv}} = \frac{-\{I_{Ai} - i_{Bi}(L - S)\}}{\{I_{Av} - i_{Bv}(L - S)\}^2} \tag{2.3.10}$$

$$\frac{\partial f_i}{\partial I_{Ai}} = \frac{1}{I_{Av} - i_{Bv}(L - S)} \tag{2.3.11}$$

$$\frac{\partial f_i}{\partial I_{Av}} = \frac{-\{I_{Ai} - i_{Bi}(L - S)\}}{\{I_{Av} - i_{Bv}(L - S)\}^2} \tag{2.3.12}$$

Thus, the error $\Delta f_i$ of the radiance of the bright boulders is given by:

$$\Delta f_i = \sqrt{\left(\frac{\partial f_i}{\partial S}\Delta \mathrm{S}\right)^2 + \left(\frac{\partial f_i}{\partial i_{Bi}}\Delta i_{Bi}\right)^2 + \left(\frac{\partial f_i}{\partial i_{Bv}}\Delta i_{Bv}\right)^2 + \left(\frac{\partial f_i}{\partial I_{Ai}}\Delta I_{Ai}\right)^2 + \left(\frac{\partial f_i}{\partial I_{Av}}\Delta I_{Av}\right)^2}. \tag{2.3.13}$$

An example of the spectrum is shown in Fig. 6A, and the breakdown of the error factors ($\frac{\partial f_i}{\partial S}\Delta \mathrm{S}$, $\frac{\partial f_i}{\partial i_{Bi}}\Delta i_{Bi}$, $\frac{\partial f_i}{\partial i_{Bv}}\Delta i_{Bv}$, $\frac{\partial f_i}{\partial I_{Ai}}\Delta I_{Ai}$ and $\frac{\partial f_i}{\partial I_{Av}}\Delta I_{Av}$) is shown in Fig. 6B. As shown in Fig. 6B, the errors attributed to the mean values of the background or shot noise ($\Delta i_{Bi}$, $\Delta i_{Bv}$, $\Delta I_{Ai}$, and $\Delta I_{Av}$) are almost the same for all band filters. The same behavior in these errors can be observed in the spectra of other bright boulders. In contrast, the errors attributed to area strongly depend on band filter. As shown in Eq. 2.3.8b, the error coefficient from the size error contains deviations between the normalized reflectance of bright boulders and the background ($I_{Ai}/I_{Av} - i_{Bi}/i_{Bv}$). As the spectra of the background (i.e., the surface of Ryugu) are almost flat, size errors correspond to the shape of the spectrum and become larger at peaks or absorptions.

We also considered the upper and lower limit of the size errors ($\Delta S_1$, $\Delta S_2$). The deviation $\Delta f_i$ of normalized reflectance with areas of $S + \Delta S_1$ and $S$ is given by

$$\Delta f_i \equiv f_i(S + \Delta S_1) - f_i(S) = \frac{I_{Ai} - i_{Bi}\{L - (S + \Delta S_1)\}}{I_{Av} - i_{Bv}\{L - (S + \Delta S_1)\}} - \frac{I_{Ai} - i_{Bi}(L - S)}{I_{Av} - i_{Bv}(L - S)} \quad . \qquad (2.3.14a)$$

$$= \frac{\Delta S_1 I_{Av} i_{Bv}}{\{I_{Av} - i_{Bv}(L - S - \Delta S_1)\}\{I_{Av} - i_{Bv}(L - S)\}} \left(\frac{I_{Ai}}{I_{Av}} - \frac{i_{Bi}}{i_{Bv}}\right). \qquad (2.3.14b)$$

The denominator of Eq. 2.3.14b is positive because $\{I_{Av} - i_{Bv}(L - S)\}$ and $\{I_{Av} - i_{Bv}(L - S - \Delta S_1)\}$ represent the total radiance of the same bright boulder. Thus, the sign of this deviation (whether this deviation is positive or negative) depends on the deviation between the normalized reflectance of bright boulders and the background. When the contrast of the bright boulder is higher than that of the background $(\text{i.e.}, I_{Ai}/I_{Av} > i_{Bi}/i_{Bv})$, this derivation $\Delta f_i$ represents the upper error of the normalized reflectance. In the opposite situation (i.e., $I_{Ai}/I_{Av} < i_{Bi}/i_{Bv}$), the deviation $\Delta f_i$ represents the lower error of the normalized reflectance. Thus, the error $\Delta S_1$ attributed to the maximum area always extends in the direction $\Delta f_i$ in which the deviation of normalized reflectance of a bright boulder and the background increases. The deviation of normalized reflectance with an area of $S - \Delta S_2$ and $S$ is simply obtained by replacing $+\Delta S_1$ with $-\Delta S_2$ in Eq. 2.3.14b and has the opposite sign.

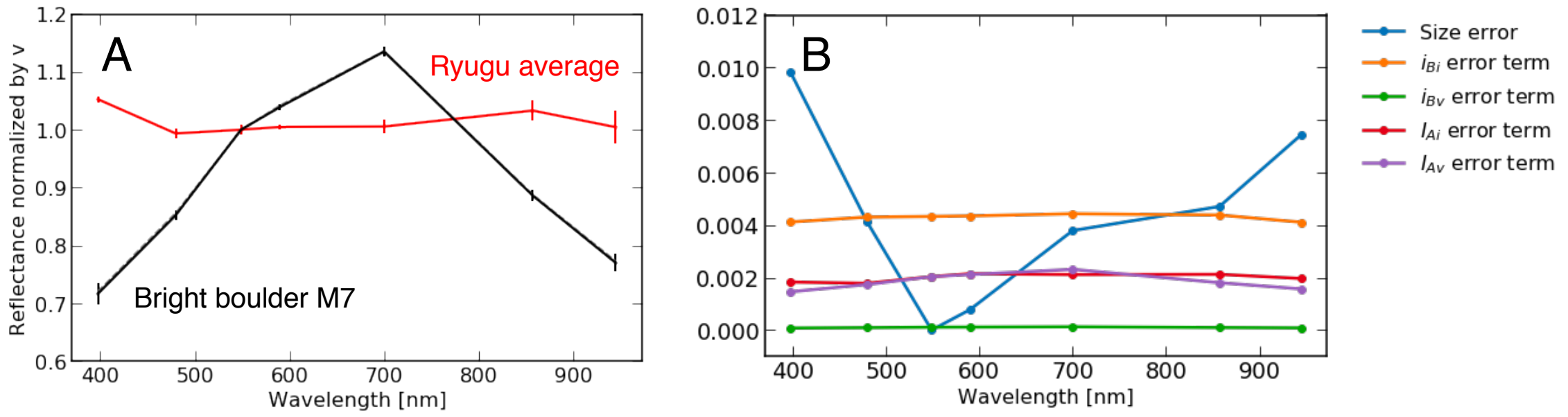

**Fig. 6** (A) Normalized reflectance spectra of the M7(C65) bright boulder (black solid line) shown in Fig. 4A. The red line is the global average of Ryugu (Sugita et al., 2019). (B) Breakdown of error factors in the error propagation for the errors shown in (A).

## 2.4 Photometric correction

The reflectance can be modeled as a function of incidence angle $i$, emission angle $e$, and phase angle $\alpha$. The I/F spectra of bright boulders extracted in Section 2.3 were photometrically corrected to the standard geometry of incidence, emission, and phase angles $(i, e, \alpha)$ = (30°, 0°, 30°), using the Hapke equations for disk-resolved reflectance (Hapke 1984). Although bright boulders and clasts are significantly brighter than Ryugu's general boulders whose (30°, 0°, 30°) reflectance factor is ~0.02 (Sugita et al. 2019; Tatsumi et al., 2020), most of them (all C-type bright boulders and many S-type bright boulders) are 1.5 to 3 times brighter; their (30°, 0°, 30°) reflectance factors are 0.03 to 0.06. Thus, we use the Hapke parameters in Sugita et al. (2019) developed for dark general boulders in this study.

Due to the rotation of the asteroid during the imaging sequence of a seven-band observation, the ROI (i.e., position of bright spot) in the FOV changes slightly over time. The effect of asteroid rotation results in a sub-degree (<1°) difference in the phase angle (i.e., Spacecraft-BB-Sun angle) between different bands. Using the pixel resolution $\delta_{pix} =$ 22.14 arcsec (Kameda et al., 2017), the line of sight (LOS) vector $\vec{R}$ from the spacecraft to a bright boulder located at $(x, y)$ in pixel coordinates is given by:

$$\vec{R} = \begin{pmatrix} tan\left\{(x-512)\cdot\delta_{pix}\right\} \\ tan\left\{(y-512)\cdot\delta_{pix}\right\} \\ 1 \end{pmatrix}, \qquad (2.4.1)$$

where $\Delta_x = x - 512$ and $\Delta_y = y - 512$ are the relative coordinates of a bright boulder from the center of the image. A schematic diagram of the LOS vector, the solar incidence vector, and the coordinate system are shown in Fig. 7. Using the solar incidence vector $\vec{S}$, the phase angle for a bright boulder located at $(x, y)$ in a pixel coordinate is given by

$$\cos\alpha = \frac{\vec{R}\cdot\vec{S}}{|\vec{R}|\,|\vec{S}|}. \qquad (2.4.2)$$

Owing to their small size, the local geometries of bright boulders cannot be derived from the global shape models of Ryugu. Thus, we conducted the photometric correction, assuming that the incidence, emission, and phase angles $(i,\ e,\ \alpha)$ are equal to ($\alpha, 0°, \alpha$), where the ROI faces the detector. However, they could have very different illumination angles. In this section, we assess the possible error from this assumption with the photometrical correction. Phase angles during the hovering observation after the MASCOT deployment, as well as the CRA1 and CRA2 observations, range from 9.9° to 14.8°, 11.3° to 29.3°, and 21.1° to 35.2°, respectively. Normal albedo $A_{normal}$ and the reflectance factor $REFF$ at ($i, e, \alpha$)=(30°, 0°, 30°) were measured as follows:

$$A_{normal} = C_0 \times RADF_{measured}(i, e, \alpha), \tag{2.4.3}$$

$$REFF(30°, 0°, 30°) = C_{30} \times \frac{RADF_{measured}(i, e, \alpha)}{\cos(30°)}, \tag{2.4.4}$$

where $C_0 \equiv ADF_{model}(0°, 0°, 0°)/RADF_{model}(\alpha, 0, \alpha)$, and $C_{30} \equiv RADF_{model}(30°, 0°, 30°)/RADF_{model}(\alpha, 0, \alpha)$.

Based on the following thought experiment, we estimated the error in the photometric correction factor $C_0$ and $C_{30}$. As the phase angle is well constrained by the position of the spacecraft, the Sun, and Ryugu, we set the incidence and emission angles as free parameters under the constraint of the phase angle and tested how much the photometric correction factor changes from those when we assumed the phase angle is equal to the incidence angle. The angles were varied under the following condition:

$$\begin{cases} |i - e| \leq \alpha \ \ (i > \alpha) \\ |\alpha - e| \leq i \ \ (i \leq \alpha) \end{cases}, \tag{2.4.5}$$

where $0° \leq i < 90°$, $0° \leq e < 90°$. We computed the variation of $C' = RADF_{model}(\alpha, 0, \alpha)/RADF_{model}(i, e, \alpha)$ when the emission and incidence angles were selected arbitrarily under the previous condition. Figure 8A and 8B show the maximum and minimum values of $C'$, respectively, when we fix the phase angles between 9° and 15° for the 2.7-km hovering observation and 11° and 36° for the 1.7-km scanning observations, respectively. This range of $C'$ values provides the errors for our estimation of albedo. Even when the incidence and emission angles are extremely high, such as 60°, the upper and lower limits are estimated as +25% and -15% for $\alpha = 15°$, and +37% and -25% for $\alpha = 36°$, respectively.

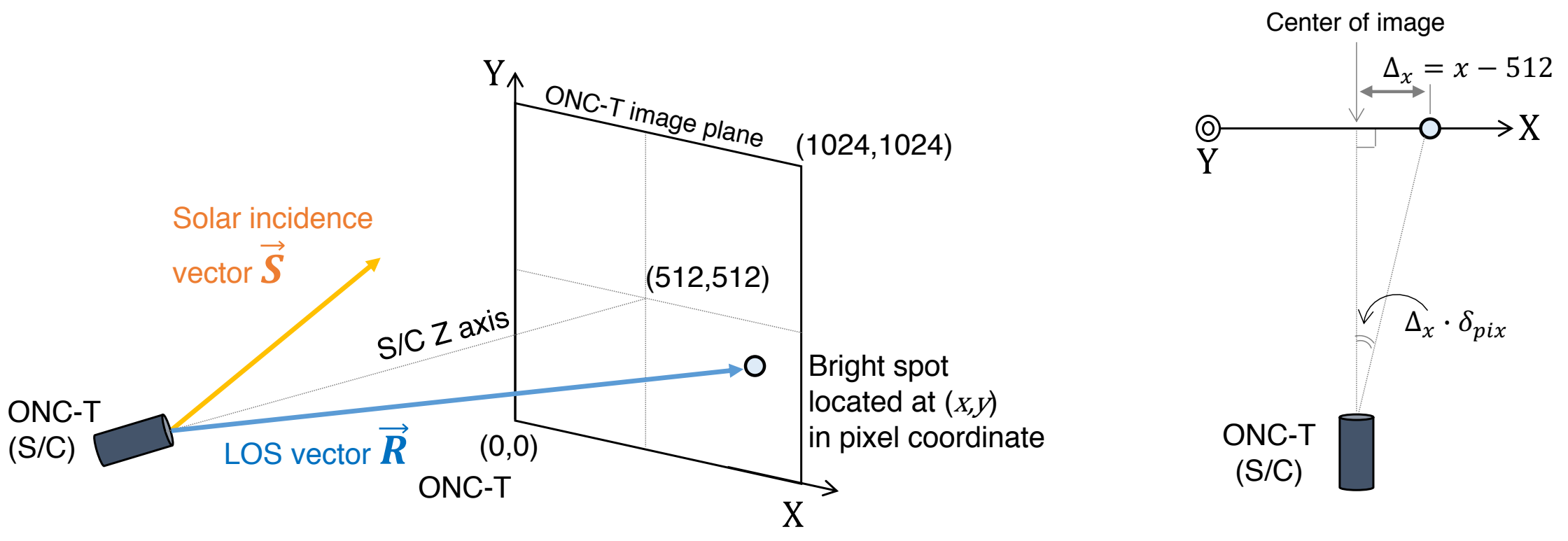

**Fig. 7** Schematic diagram of line-of-sight vector and solar incidence vector (left) and a top view of the image plane coordinate system (right).

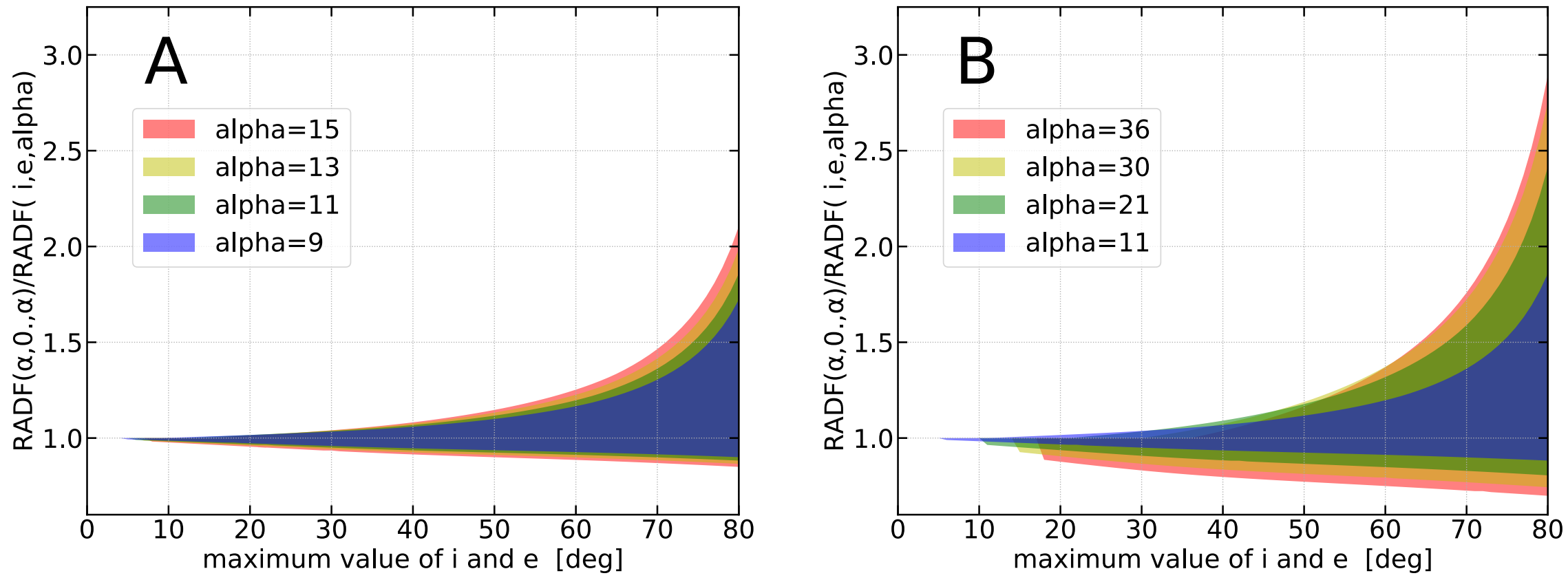

**Fig. 8** Possible values of the photometric correction factor $C'$ when the phase angle α is set to (A) 9°, 11°, 13°, and 15° and (B) 11°, 21°, 30°, and 36°. Different colors indicate the range of $C'$ for incidence and emission angles permitted for phase angles $\alpha$ indicated in the figure under the condition of Eq. 2.4.5.

## 3. Analysis Results

The analysis conducted in this study revealed a number of important spectral properties of bright boulders on Ryugu. In the following section, we first discuss the spectral classification of bright boulders newly observed during the 1.7-km scanning operations (~19 cm/pixel).

Higher-resolution images analyzed in this study allowed us to obtain more accurate reflectance measurements of small bright boulders than that of Tatsumi et al. (2021). During the 1.7-km scanning 4-band observations (CRA1 and CRA2), 79 bright boulders (C1–C79) were observed, among which 76 bright boulders newly found by this study. Note that most (18) of the bright boulders found by Tatsumi et al. (2021) were not covered in the CRA1 and CRA2 observations. Three bright boulders (i.e., C65/M7, C29/M19, and C26/M21) observed in the images of the 1.7-km scanning observations are the same bright boulders observed during the 2.7-km hovering observation and analyzed in Tatsumi et al. (2021).

Four-band (ul, b, v, and x bands) spectra of all these 79 bright boulders were obtained in this study (Fig. 9). The analysis results show that most of the newly observed bright boulders exhibit featureless flat spectra and that only a small fraction of them exhibit spectra that greatly deviated from the global average of Ryugu. Visual inspection of the 79 spectra shows that some of these bright boulders indicate strong absorption at wavelengths shortward of 0.55 $\mu$m (i.e., low ul/v ratio); the others show almost flat spectra. As we only have 4-band coverage here, lacking information on the 1-$\mu$m absorption and the 0.7-$\mu$m peak (i.e., p/w ratio), low ul/v ratio is not strong evidence for S-type spectra. However, it is consistent with an S-type spectrum. Furthermore, the spectra with deep UV absorption are sometimes accompanied by a blue v-to-x slope.

To address this problem, we conducted a more quantitative screening using the ul/v ratio derived from the 0.19-m/pixel resolution images. More specifically, we selected a ul/v ratio of 0.87, which is the middle point between the S-type Main-Belt Asteroid (MBA) and C-type MBA spectra, as the threshold for S-type candidate bright boulders. All the S-type bright boulders found by Tatsumi et al. (2021) have a ul/v ratio $< 0.87$. Hence, eight bright boulders (i.e., C4, C6, C25, C26, C28, C51, C56, and C65) meet this criterion for S-type candidates. Then, we evaluated the seven-band spectra of these bright boulders using the following method.

The sizes of these boulders were measured from v-band images obtained at 1.7 km. Using these boulder sizes, we evaluated the spectra of four newly found S-type candidate bright boulders (i.e., C4, C25, C28, and C56) using seven-band images taken during 2.7-km hovering observation. Note that C6 and C51 were not observed during the 2.7-km hovering observation because they are located at higher latitudes, which are outside the FOV. Thus, we could not clarify the spectral types of C6 or C51. Furthermore, because bright boulders C26 (M21) and C65 (M7) are classified as C and S types, respectively, by Tatsumi et al. (2021), further analysis is not necessary.

Figure 10 presents the comparison of spectra derived from seven-band images and those derived from four-band images, showing that these spectra agree with each other within errors. Out of eight S-type candidate bright boulders, two bright boulders exhibit diagnostic features for the S-type spectra. The 7-band spectrum of C25 shows an apparent reflectance peak at ~0.7 μm and absorption at ~1 μm, suggesting that this is an S-type bright boulder. That of C4 also shows a weak but clear reflectance peak at ~0.7 μm and absorption at ~1 μm, suggesting that this is also an S type. Its error at the ul band is very large, but this might be due to the shadows of adjacent boulders. Bright boulder C4 exhibits similar spectra regardless of the sizes of $ROI_A$ and $ROI_B$. The same is true for the other S-type-candidate bright boulders shown in Fig. 10. Thus, we classify C4 and C25 as S-type bright boulders. Noted here is that the fact that these boulders exhibit S-type spectra indicates that it does not have a large amount of serpentine, which would show 0.7 μm-absorption. Thus, these boulders did not experience extensive aqueous alteration. Furthermore, it is also noted that one C65 (M7) of these S-type bright boulders is shown as a clast embedded in the large darker substrate boulder as described in Paper1, which indicates that fragmentation and cementation of these boulders occurred after the termination of aqueous alteration.

In contrast, the spectra of C56 and C28 show no apparent reflectance peak at ~0.7 μm but have a strong UV absorption and red spectral slope, which is consistent with hydrated carbonaceous chondrites. Laboratory experiments suggest that heating of carbonaceous chondrites decreases their UV absorption and makes the spectral slope flatter (e.g., Hiroi et al. 1996a; Hiroi et al. 1996b; Cloutis et al., 2012). The spectra of two heated carbonaceous chondrites are shown in Fig. 11. The spectra of C56 have a UV absorption of ~20% and a flat v-to-p slope, which is consistent with the spectra of Ivuna heated at 100–500 °C and Murchison heated to ~600 °C (Hiroi et al., 1996a; Hiroi et al., 1996b). In contrast, the spectrum of C28 exhibits a relatively strong UV absorption of ~40% and has a possible absorption at ~0.7 $\mu$m, which is consistent with unheated Murchison and implies the presence of an Fe-bearing phyllosilicate (Vilas and Gaffey, 1994). Thus, we conclude that C56 and C28 are C-type bright boulders.

The reflectance factor at incidence, emission, and phase angles ($i, e, \alpha$) = (30°, 0°, 30°) of bright boulders observed during the 1.7-km scanning observations are shown in Fig. 12. As absolute reflectance is extremely sensitive to the sizes of bright boulders, bright boulders covering areas < 10 pixel$^2$ (Fig. 12B) have larger errors than bright boulders covering areas ≧ 10 pixel$^2$ (Fig. 12A). Comparison of bright boulders with areas ≧ 10 pixel$^2$ suggests that S-type bright boulder (C65) is likely to have a higher reflectance factor than C-type bright boulders, similar to the analysis using lower resolution images by Tatsumi et al. (2021). More specifically, although the large error bar of the reflectance factor of C65 overlaps with the brightest C-type bright boulders, the reflectance factor of S-type bright boulder (C65) under the zero-emission-angle ($e$ = 0°) assumption is higher than the average (~3%) of C-types under the same assumption by more than $2\sigma$ of their distribution, which is consistent with the results of bright boulders with areas ≧ 10 pixel$^2$ in lower resolution images by Tatsumi et al. (2021). Furthermore, the reflectance of this S-type bright boulder (C65) measured in the 1.7-km scanning observation images is ~5%, which is consistent with the reflectance of ~5% derived from other datasets (i.e., the 2.7-km hovering observation images). In contrast, most of the bright boulders with areas < 10 pixel$^2$ have very large errors (~ 14 %). Thus, the detailed spectral properties of these small bright boulders are difficult to evaluate. Nevertheless, the newly found S-type bright boulders (i.e., C4 and C25) have relatively small errors, and their reflectance of ~5% is consistent with other S-type bright boulders.

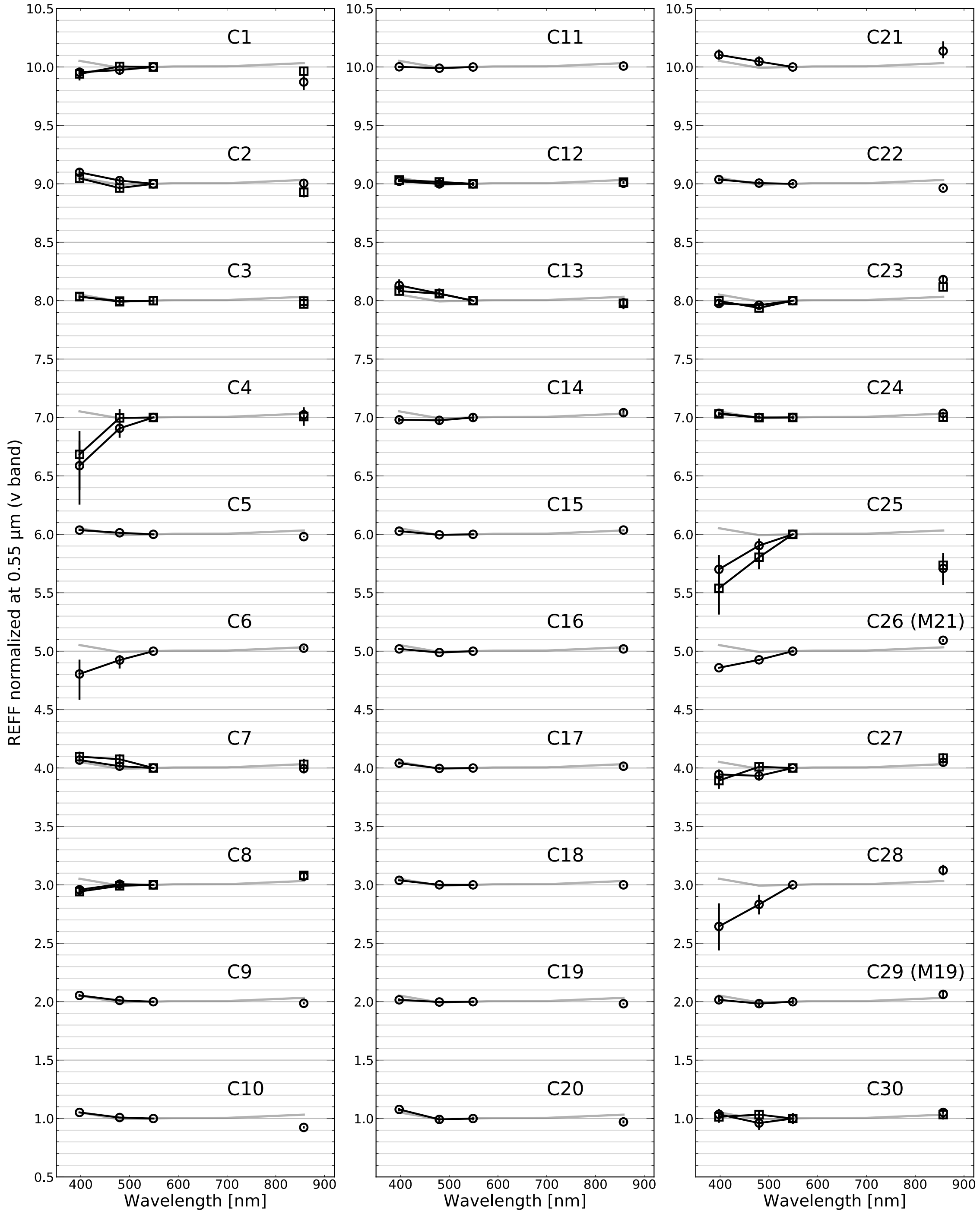

**Fig. 9** Relative reflectance spectra of bright boulders observed during the 1.7-km scanning observations (CRA1 and CRA2). The spectra are normalized at 0.55 μm. Bright boulders (black lines) are compared with the average spectrum of Ryugu (gray lines) by Sugita et al. (2019). Lines are offset by unity for clarity, and bright boulders with several spectra are observed multiple times. In order to distinguish data points in different spectra, we use different symbols (squares, circles, and triangles) in the figure. Solid lines are used

to connect adjacent filter bands, whereas dashed lines are used to connect nonadjacent bands.

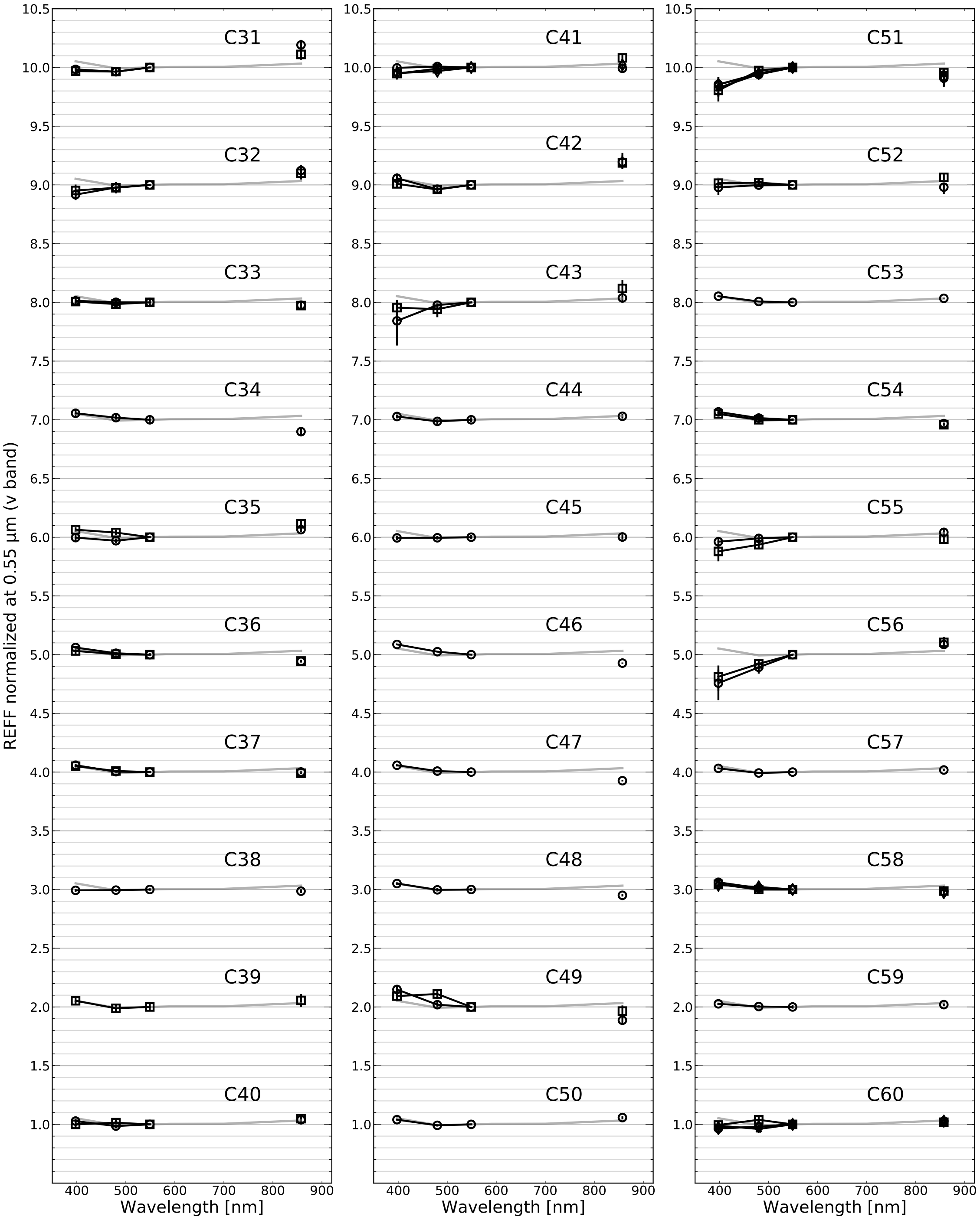

**Fig. 9 (Continued)**

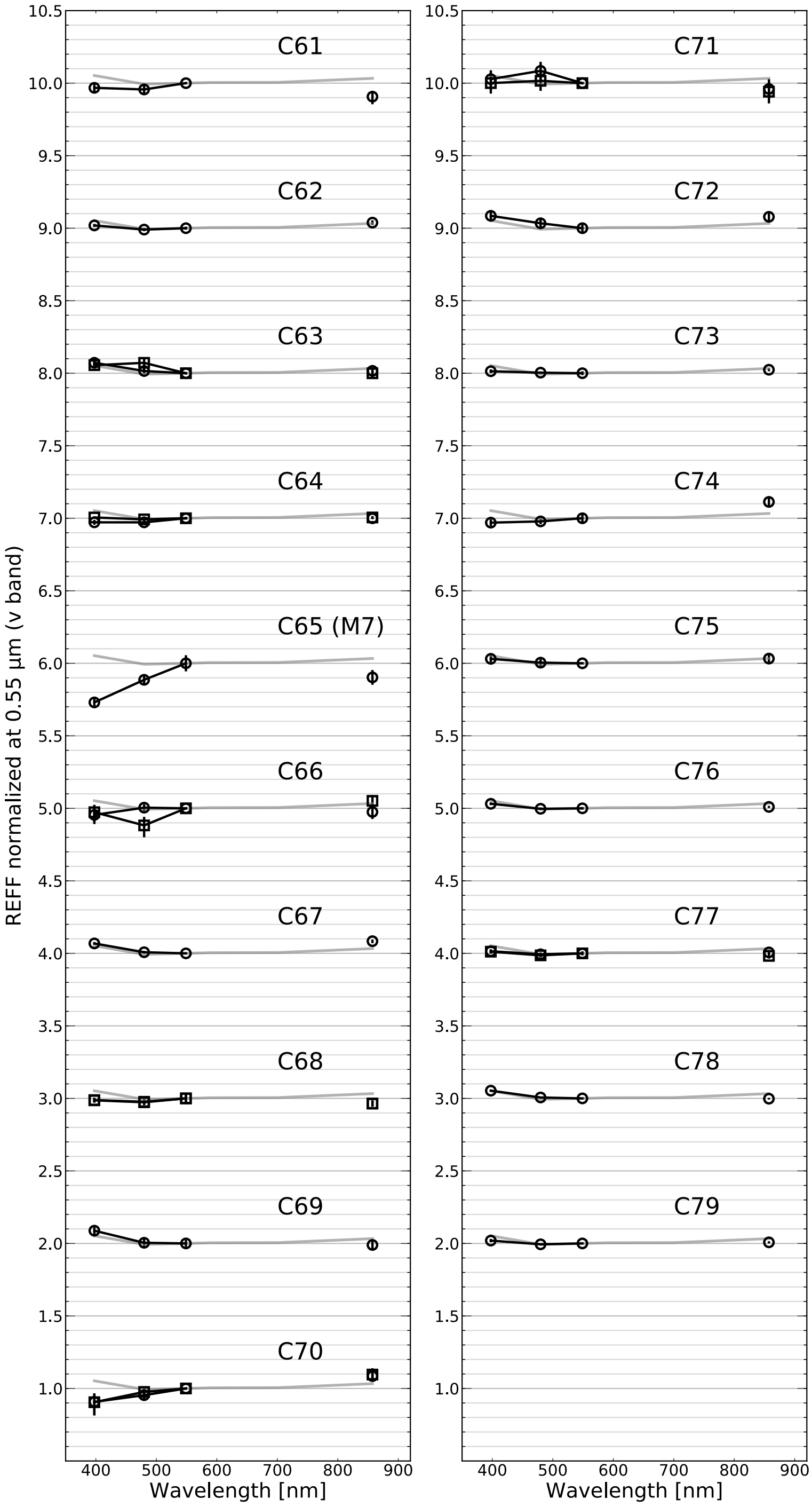

**Fig. 9 (Continued)**

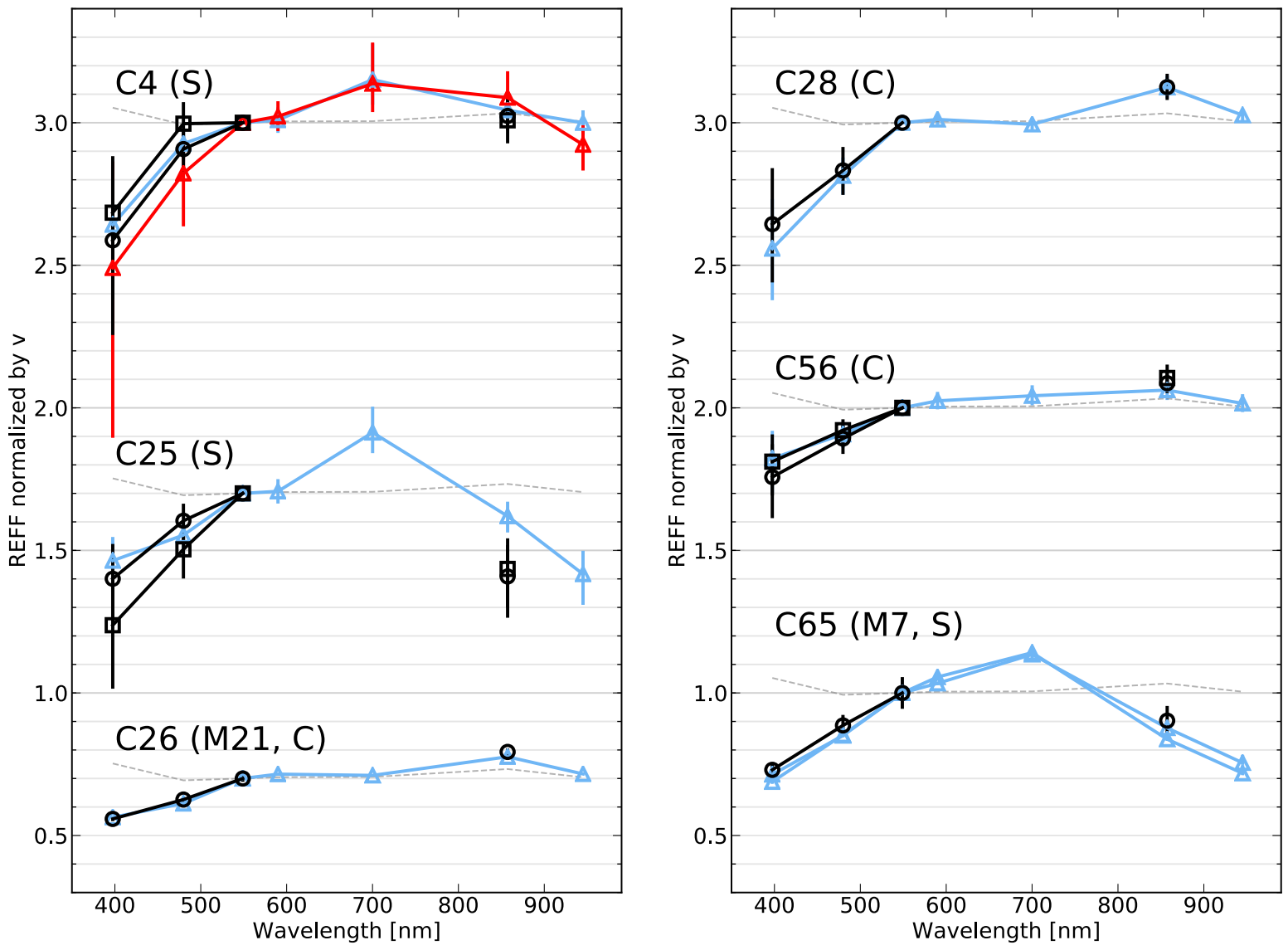

**Fig. 10** Normalized reflectance spectra of S-type-candidate bright boulders observed during the 1.7-km scanning (black lines) and the 2.7-km hovering operation (light blue and red lines). The light blue lines indicate the spectra derived using larger $ROI_A$ and $ROI_B$ (side length of 5 and 15 pixels, respectively). The red lines indicate the spectrum derived using smaller $ROI_A$ and $ROI_B$ (side length of 3 and 9 pixels, respectively). Gray dashed lines are the average spectrum of Ryugu (Sugita et al. 2019), and all the spectra are normalized at 0.55 $\mu$m and offset from each other vertically by unity for clarity. Our spectral type classification is indicated in the parentheses.

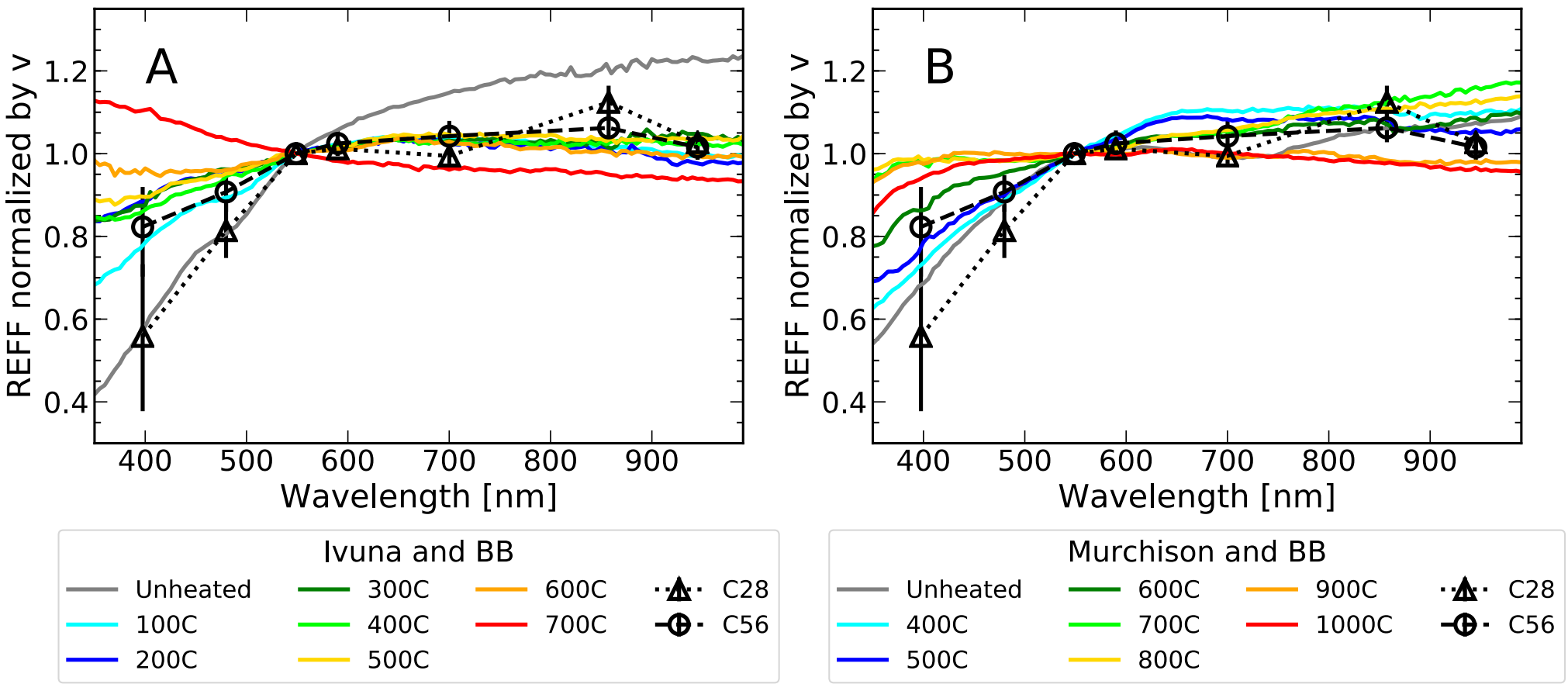

**Fig. 11** (A) Normalized reflectance spectra of heated Ivuna samples and bright boulders (C28, C56). (B) Normalized reflectance spectra of heated Murchison samples. The data are from Hiroi et al. (1996a) and Hiroi et al. (1996b), as well as bright boulders (C28,

C56). The colors represent different temperatures indicated in the bottom box. BB stands for Bright Boulder.

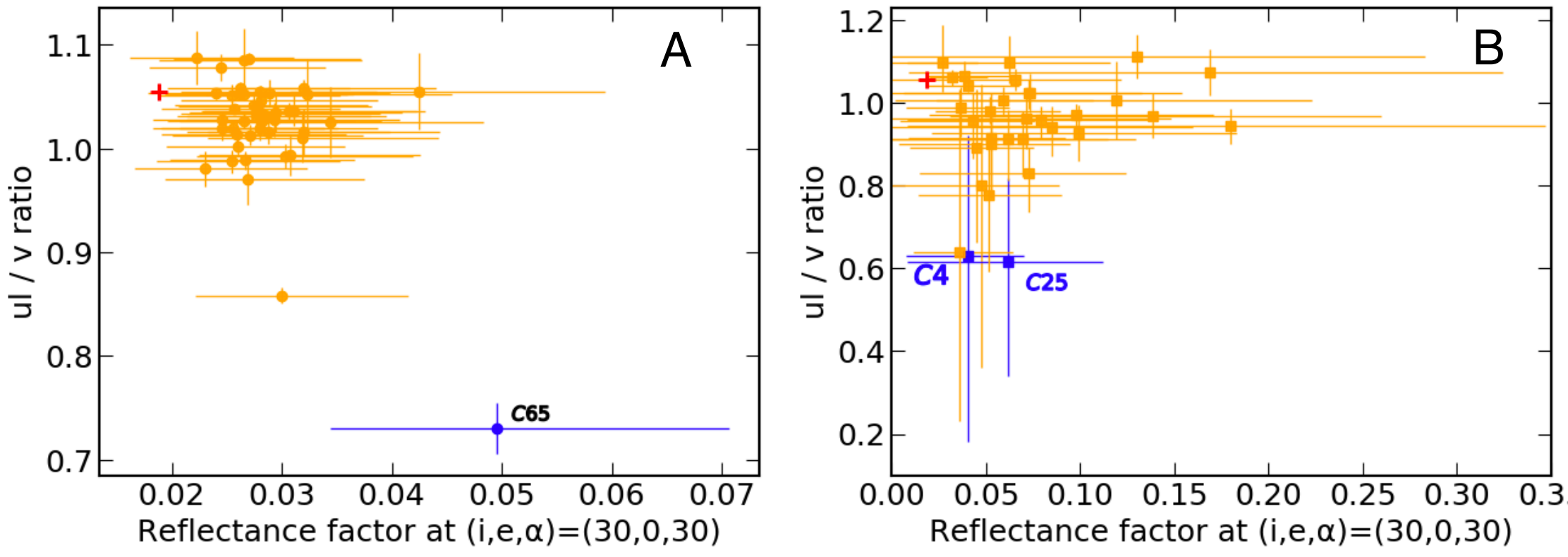

**Fig. 12** V-band (0.55 $\mu$m) reflectance factor at incidence, emission, phase angles $(i, e, \alpha)$=(30°, 0°, 30°), and ul/v band ratio of bright boulders. (A) Orange (C-type) and blue (S-type) circles are bright boulders with areas ≧ 10 pixel$^2$. (B) Orange (C-type) and blue (S-type) squares are bright boulders with areas < 10 pixel$^2$. The red crosses indicate the global average of Ryugu (Sugita et al. 2019).

## 4. Discussion

The analyses of bright boulders performed in this study have various implications for the evolution of the asteroid Ryugu. In this section, we discuss these implications through comparisons with meteorite spectra, laboratory experiments, and main-belt asteroids. More specifically, we discuss the spectral variety among S-type bright boulders in Section 4.1, that among C-type bright boulders in Section 4.2, and constraints on the evolution of Ryugu and its parent body in Section 4.3.

### 4.1 Spectral variety of S-type bright boulders

The presence of S-type bright clasts embedded and mechanically adhered in dark boulders (Section 4 in Paper 1) is consistent with a hypothesis; that is, S-type bright boulders were mixed before or during Ryugu's formation event via a catastrophic disruption impact on Ryugu's parent body (Morota et al., 2020), as proposed by Tatsumi et al. (2021). Furthermore, the spectral variety among S-type bright boulders further constrains the collisional history of Ryugu and its parent body. The visible spectra of S-type bright boulders show variation in both spectral slope and absorption at ~1 $\mu$m. If their spectral variety is large enough to require multiple meteoritic materials, it would suggest that they are from multiple projectiles. In contrast, if their spectral variety is consistent with spectra derived from a single starting material, it would instead support that they came from a single impactor. Here, we assume a single S-type projectile containing a single meteoritic type. Although multiple ordinary chondrite compositions can occur on a single S-type asteroid in theory, disk-resolved spacecraft observations of S-type asteroids have not provided evidence to support such a possibility, thus far. For example, the majority of the observed spectral variety on Itokawa's surface can be accounted for by variations in maturity due to space weathering and do not require compositional variation (Koga et al., 2018). Furthermore, grain size variation and dust cover also could affect the spectra of S-type bright boulders.

To examine whether these spectral modification effects can account for the observed spectral variety of S-type bright boulders on Ryugu or compositional variation is required, we compared the spectra of S-type bright boulders with both main-belt S-type asteroids and meteorite samples that have been subjected to different processing (e.g., space-weathering experiments and size sorting). We conducted principal component analysis (PCA) using the data collected in the second phase of the Small Main-belt Asteroid Spectroscopic Survey (SMASSII) by Bus and Binzel (2002) and compared the spectra of the S-type bright boulders with those of meteorites. To examine the effect of space weathering, compositional difference, and grain size on the visible spectra of ordinary chondrites, we used the spectral data of ordinary chondrites exposed to pulsed-laser irradiation (to simulate space weathering), considering also various sub-types, and different grain sizes cataloged in Reflectance Experiment LABoratory (RELAB) of Brown University. To examine the effect of dust cover, we calculated the linear combinations of the spectra of bright boulders and chondrites.

We conducted PCA, based on the method described by Sugita et al. (2019), using the SMASSII data (Bus and Binzel, 2002) (Fig. 13). The first principal component, PC1, and the second component, PC2, represent the spectral slope shortward of 0.70 $\mu$m and the 1-$\mu$m absorption, respectively (Fig. 13B). Thus, larger PC1 and PC2 score values correspond to a redder spectral slope and a deeper absorption at ~1 $\mu$m, respectively. We examined the spectral variety of six S-type bright boulders (i.e., M7, M8, M9, M13, M16, and M20) found by Tatsumi et al. (2021) and two additional S-type bright boulders (i.e., C4 and C25) shown in Section 3 from this study.

In the following sections, we first describe the spectral variety of S-type bright boulders in comparison with MBAs from SMASSII. Then, we discuss whether the spectral variety in the S-type bright boulders can be explained by possible processes, such as space weathering, compositional variation, thermal metamorphism, grain size variation, and dust cover by comparing meteorite spectral data from RELAB.

*Comparison with MBAs*: The PC2 values (i.e., 1-$\mu$m absorption depth) of the eight S-type bright boulders range from S-, Q-, O- to V-type MBAs (Fig. 13C). Although the error bars for some bright boulder spectra are large, their distribution range is much larger than these uncertainties. The range of PC1 values (i.e., spectral slope) indicates that the S-type bright boulders tend to be bluer than S- and V-type MBAs. Furthermore, a clear correlation between spectral slopes (PC1 values) and 1-$\mu$m absorptions (PC2 values) is observed for S-type bright boulders; bluer bright boulders tend to have a deeper 1-$\mu$m absorption. This trend is consistent with the space weathering of mafic minerals; space weathering reddens the general slope and decreases the 1-$\mu$m absorption (e.g., Sasaki et al., 2001; Brunetto et al., 2015). This trend further coincides with a transition in spectral properties from Q- and Sq- to S-type MBAs, similar to that found on the surface of the S-type asteroid Itokawa (Koga et al., 2018). Thus, this spectral variation observed among the S-type bright boulders on Ryugu likely reflects differences in exposure age.

*Comparison with laser-irradiated ordinary chondrites*: To examine the above possibility that the spectral trends in S-type bright boulders properties observed in the PC-space comes from space weathering, we compared the spectra of ten ordinary chondrites processed by pulsed laser irradiation (RELAB) with the bright boulder spectra (Fig. 13C and Fig. 14). Laser irradiation has been used for simulating space weathering due to micrometeorite bombardment. Thin surface rims containing nanophase-Fe very similar to that seen in the returned samples from the Moon and Itokawa have been reproduced in these laboratory experiments (Sasaki et al., 2001; Noguchi et al., 2011; Brunetto et al., 2015). Details of the chondrites used in this study (e.g., sample name, grain size, and the energies of the laser irradiation applied to them) are described in Appendix (Table A.1).

Comparison between S-type bright boulders on Ryugu and laser-irradiated chondrites reveals a great similarity between the irradiation trends and the distribution of S-type bright boulders on Ryugu. Almost all these ordinary chondrites redden as the laser energy increases. They further exhibit almost the same slope, $a \sim -0.52$, in the PC space. Here, the PCA irradiation trend values are related by $\mathrm{PC2} = a \times \mathrm{PC1} + b$, where $b$ is the PC2 intercept. The direction of changes in the spectral characteristics of the chondrite spectra induced by laser irradiation coincides with the spectral trend observed for the S-type bright boulders, whose slope in the PC space is $a \sim -0.55$. Furthermore, the range of spectral slopes for the S-type bright boulders is also comparable to that of the laser-irradiated chondrites. These similarities between Ryugu S-type bright boulders, laser-irradiated ordinary chondrites, and MBAs, strongly suggests that the variation observed along the PC1–PC2 trend among S-type bright boulders reflects different maturity levels due to space weathering.

In contrast, the variation in the absolute value of the 1-$\mu$m absorption seen in S-type bright boulders is less likely due to space weathering. Observed S-type bright boulders appear to follow two trends consistent with space-weathering tracks: trend I starting with initial spectrum with shallower 1-$\mu$m absorption and trend II starting with initial spectrum with a deeper absorption. The difference in PC2 intercept $b$ of these two trend lines is ~ 0.16, which is larger than the PC2 errors of individual S-type bright boulders (Fig. 15A and B).

Using the method described in Koga et al. (2018) and some meteorite samples, we estimated the exposure times of these S-type bright boulders on Ryugu. Such age estimations for the Ryugu surface may have very important implications for small body cratering efficiency because the surface age interpretations may be more complicated than previously thought. For example, a recent investigation on the surface age and dynamic evolution for spacecraft-explored asteroids, including Ryugu, found a perplexing relation between crater retention age and formation age (Bottke et al., 2020). Ballouz et al. (2020) found many "mini-craters" (decimeter to meter) on Bennu's boulders and estimated young boulder surface ages (~$2 \times 10^6$ years).

Since the number of nuclear displacements, $d$, per unit surface area is proportioned to the ion fluence (Brunetto and Strazzulla, 2005), the relationship between the irradiation time scale $t$ (yr) and nuclear displacement $d$ ($\mathrm{cm}^{-2}$) is given by:

$$t = 2.53 \times 10^{-13} \times \left(\frac{D}{2.9\ \mathrm{AU}}\right)^2 d, \tag{4.1.1}$$

where $D$ is the distance from the sun (Brunetto et al., 2006). Considering that Ryugu is at ~1.2 AU from the Sun, the irradiation time scale $t$ can be obtained by $t \sim 4 \times 10^5 (d/10^{19})$ yr. The relationship between irradiation energy and nuclear displacements shown in Koga et al. (2018) suggests that laser energies of 15–80 mJ corresponding to nuclear displacements of $d \sim 10^{19}$ to $6 \times 10^{19}$ cm$^{-2}$. Thus, comparison with laser irradiation experiments using ordinary chondrites leads us to estimate that the irradiation timescale of S-type bright boulders on Ryugu would range from $10^5$ to $10^6$ years (Fig. 14). These timescales are consistent with the average resurfacing age (<$10^6$ yr) of the top ~1-meter layer on Ryugu (Sugita et al., 2019; Arakawa et al., 2020; Morota et al., 2020). Furthermore, the presence of bright boulders in the interior of the SCI crater, which is described in Section 4 in Paper I, indicating that bright boulders exist not only on the surface of Ryugu but also in the near subsurface. This implies that S-type bright boulders may be transferred from the subsurface to the surface by mass flow processes, such as cratering, seismic shaking, or landslides on Ryugu.

*Comparison with ordinary chondrites with different compositions*: As laser irradiation trends are almost parallel to each other, their difference in PC2 intercepts $b$ (i.e., absorption at ~1 $\mu$m) depends on their initial pre-irradiation spectra. The differences in the composition of the starting materials may be the cause for the differences in spectral properties of unirradiated ordinary chondrites. Thus, we examined whether different compositions (e.g., H, L, and LL) or the degrees of metamorphism (3–6) of ordinary chondrites may correlate with the PC values of their spectra.

Comparison among different groups of ordinary chondrites with the same sample preparation (i.e., chip) shows that spectral properties do not have a clear systematic difference in the wavelength range (0.40–0.95 $\mu$m) covered by ONC-T filters (Fig. 16). Thus, the observed variations in spectral slope and 1-$\mu$m absorption depth among ordinary chondrites may reflect scatter among individual meteorite samples, rather than their composition or metamorphism. This result is consistent with analysis by Vernazza et al. (2008) that H-, L- and LL-chondrite meteorites are difficult to distinguish at wavelengths shorter than ~1 $\mu$m. Nevertheless, if two projectiles analogous to different groups of ordinary chondrites hit Ryugu's parent body(s), two distinctively different spectral trends correlating to variable degrees of space weathering would be observed in the PC analysis of the S-type bright boulder spectra.

*Grain size effect*: Although boulder size can be measured from images, various surface physical properties, such as roughness or porosity, cannot be readily estimated from the imaging data. Here, we examine grain size effects as an example of such surface physical properties. Laboratory experiments suggest that decrease in particle size results in decreases in absorption depth and redder continuous slope for ordinary chondrites (Fenale et al., 1992; Clark et al., 1992). Thus, grain size effects could contribute to the difference between the apparent two PC1–PC2 trend lines observed within the S-type bright boulders on Ryugu.

The degree of this effect could be large enough to account for the observed spectral variety in S-type bright boulders on Ryugu. For example, the width of the PC2-range of Saratov (L4) with different grain sizes is $\sim 0.15$ (Fig. 13C). This is comparable to the difference in the PC2 intercept $b \sim 0.16$ of the two trend lines observed for the S-type bright boulders on Ryugu. However, the effect of grain size cannot readily explain the apparent double peak distribution of the PC2 intercept (i.e., space-weathering corrected 1-$\mu$m absorption depth) found among S-type bright boulders on Ryugu.

*Dust/regolith covering effect*: During the touchdown operation of the Hayabusa2 spacecraft on 21 February 2019, dark fine grains < 300 $\mu$m in diameter were attached to the front filter of the nadir-viewing wide-angle optical navigation camera (ONC-W1) (Morota et al., 2020). In addition, global imaging observations have revealed that many boulders on Ryugu are covered with regolith (Sugita et al., 2019). Photometric studies of the surface also indicate the presence of a fine-grained component in the regolith (Yokota et al. 2020; Domingue et al. 2020). Such regolith emplacement indicates a possible dust coating on fresh boulders, which can affect their spectral properties. As the global average spectra of Ryugu is flat in the visible region, contamination by dust or regolith of Ryugu may weaken the absorption features, such as the 1-$\mu$m absorption, of S-type bright boulders. Different degrees of coating by dust/regolith may account for the differences in the 1-$\mu$m absorption between the two trend lines for the S-type bright boulders. Furthermore, such spectral modification may also occur if S-type bright clasts occur within large breccias on Ryugu as shown in Paper 1. Then, resulting spectra of such breccia containing S-type clasts would be expressed with a linear combinations with various mixing ration. Thus, to examine the effect of dust coating on bright boulders and mixing within breccias, we calculated linear combinations with various mixing ratios, using the normalized spectra of S-type bright boulders (M7, M9, M16, and C25) with a deeper 1-µm absorption and the global average spectrum of Ryugu.

The results of this examination showed several disagreements with the observed properties of the S-type bright boulders. First, the calculation results indicate that a large fraction (50–75%) of the surface of these bright boulders need to be covered by the dust of Ryugu's average spectrum to make their absorption band as shallow as the other S-type bright boulders (M8, M13, M20, and C4) (Fig. 17A). When bright boulders are covered with the dust from Ryugu, their other spectral properties are also expected to become closer to the global average spectrum of Ryugu. However, the variations in spectral slope and reflectance of S-type bright boulders exhibit no consistent trend in this direction (Fig. 17B). Second, if the differences in color of the two trend lines are due to the coating of the dust from Ryugu, the group along the trend line with shallower 1-µm absorption should have a narrower range in spectral slope because all the bright boulders in this group have a spectrum close to the Ryugu average. However, the two trends exhibit almost the same range in spectral slope (Fig. 17A). Third, the reflectance of the two most well-resolved bright boulders (M7 and M13) suggests an inconsistent trend: bright boulders with a shallower absorption exhibit brighter reflectance than the others (Fig. 17B). These inconsistencies suggest that the two distinct trends seen among the S-type bright boulders are unlikely originated from the different degrees of dust coverage or mixing ratios of S-type clasts with breccias. This conclusion is consistent with the in-situ observations by the MASCOT lander that the boulder surfaces do not appear to be covered with fine-grain dust layers (Jaumann et al., 2019). Moreover, the thermal inertia of boulders is inconsistent with a fine-grain dust coating (Grott et al., 2019).

*Comparison with bright boulders on Bennu*: DellaGiustina et al. (2021) show that asteroid Bennu also has bright boulders with spectra consistent with anhydrous silicates, strongly suggesting an exogenic origin. Separate studies of the bright boulders on Bennu and Ryugu, both of which have been estimated to originate from the inner main belt (Campins et al., 2010, 2013, Bottke et al., 2015), suggest that they have different spectral properties, supporting collisions with different projectiles (Tatsumi et al., 2021, DellaGiustina et al., 2021). More specifically, many of Bennu's anhydrous bright boulders have a deep absorption at 1 and 2 $\mu$m (indicative of pyroxene), which similarity to V-type MBAs. In contrast, many of Ryugu's bright boulders have a shallow 1-$\mu$m absorption, and a 2-$\mu$m absorption has not been detected. However, their spectra have not yet been compared directly. As the OSIRIS-REx Camera Suite (OCAMS) filter bands (b': 0.47 $\mu$m, v: 0.55 $\mu$m, w: 0.70 $\mu$m, x: 0.85 $\mu$m) (Rizk et al., 2018) are very similar to ONC-T, we can conduct PCA using their visible spectral data.

Fig. 13 presents the PCA results, indicating that the general distribution of Bennu's bright boulders and that on Ryugu have both similarity and difference. One important similarity is that bright boulders on both Bennu and Ryugu are bluer than S-type and V-type asteroids, indicating that the exposure ages of these bright boulders are likely shorter than those of MBAs. The difference is the specific locations of the bright boulders in the PC space. We have two interpretations for the differences in their distributions. (1) Bennu's anhydrous bright boulders follow a single trend with a slope different from that of Ryugu's S-type bright boulders in PC1–PC2 space. (2) Bennu's anhydrous bright boulders follow three trends with a slope approximately the same as that of S-type bright boulders on Ryugu in PC1–PC2 space. The first interpretation would be possible if Bennu's V-type bright boulders have experienced a space-weathering trend different from the boulders on Ryugu. We considered this possibility, as the anhydrous silicate bright boulders on the two asteroids may have different compositions (HED-like vs ordinary-chondrite-like materials), which may, in turn, respond differently to space weathering processes. Accordingly, we analyzed the spectra of laser-irradiated HED meteorites from the RELAB database and the work of Hiroi et al. (2013). Although the amount of data available is small (only 3 samples and 17 spectra) and one spectrum appears to deviate from the other spectra, HED meteorites exhibit generally similar spectral trends as ordinary chondrites due to space-weathering (Fig. 13C). Note that some materials on Vesta exhibit bluing upon space weathering rather than reddening as seen on Moon, Itokawa, or other S-type asteroids (Pieters et al., 2012). However, the observed bluing trend on Vesta is not consistent with the spectral distribution among exotic boulders on Bennu by DellaGiustina et al. (2021). Thus, we infer that different responses to space weathering unlikely lead to separate PC1–PC2 trends for the anhydrous silicate bright boulders on Bennu and Ryugu.

Therefore, the second interpretation (i.e., three groups of anhydrous bright boulders on Bennu) becomes more plausible. Such possible clustering might be more easily observed in the large scale PC1–PC2 diagram (Fig. 15A) and in the distribution of PC2 intercept $b$ values (Fig. 15B). These diagrams show: (1) Most (5 out of 7) bright boulders on Bennu are relatively close to trend II on Ryugu, and (2) one bright boulder on Bennu shows a much larger intercept ($b$ ~ 0.3) than either trend within the Ryugu bright boulders. In contrast, only one bright boulder found on Bennu thus far is close to trend I on Ryugu. Furthermore, the degrees of overlap between the different types of meteorites and anhydrous bright boulders on Ryugu and Bennu vary substantially (Figs. 15B and C). More specifically, Ryugu's trend I overlaps with ordinary chondrites very well but only barely with the HEDs. Only 6.1% of HEDs exhibit $b$ values the same (0.04) or lower than trend I, whereas 28.8% of ordinary chondrites exhibit $b$ values the same or lower than trend I. Bennu's anhydrous bright boulders with larger intercepts ($b$ > 0.32) show the opposite correlations. Only 3.2% of ordinary chondrites have $b$ values the same (0.32) or higher than these anhydrous bright boulders, whereas 11.4% of HEDs have $b$ values the same or higher. This contrasting difference strongly supports that bright boulders on both asteroids came from two different projectiles. The situation of Ryugu's trend II or Bennu's bright boulders with medium $b$ values (0.12–0.22) is more complicated. As their $b$ values overlap with both HEDs and ordinary chondrites very well, we cannot tell which type of projectile is more likely, solely based on $b$ values. These observations lead us to infer that Ryugu's parent body(s) might have been hit by two projectiles and that at least one is an S-type asteroid (i.e., ordinary chondrite) and that Bennu's parent body might also have been hit by multiple projectiles, including at least one V-type asteroid (i.e., HED meteorite). Whether the other projectile for Ryugu (i.e., trend II) is the same as that for Bennu is difficult to tell yet. Regardless of specific uncertainties, the overall trend of Bennu's bright boulders in PC1–PC2 space shows a pattern different from that of Ryugu, suggesting that projectiles for the two asteroids may be different. These observations support that Ryugu and Bennu have experienced different collisional histories as suggested by Tatsumi et al. (2021) and DellaGuistina et al. (2021) and perhaps had different intermediate parent bodies.

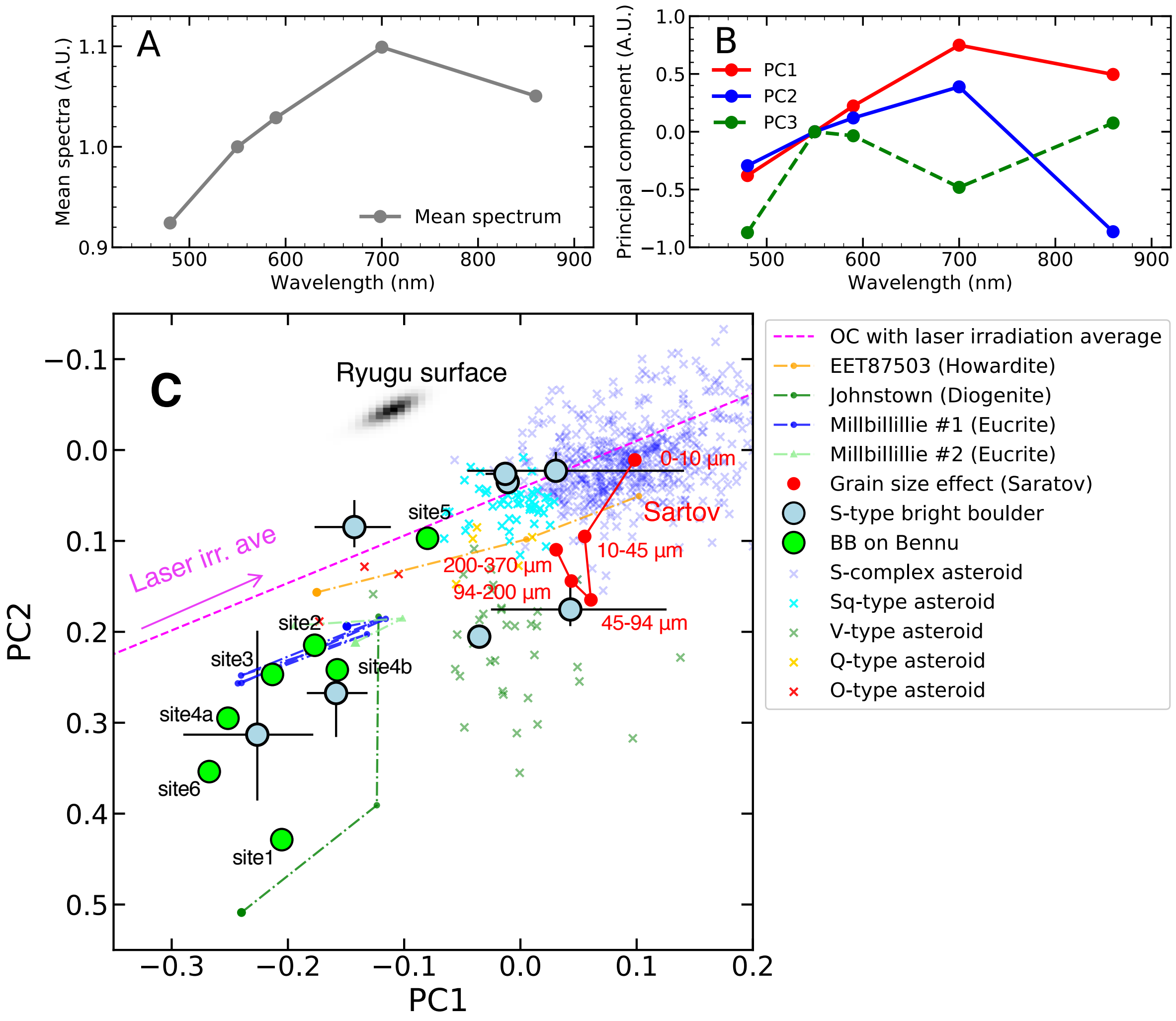

**Fig. 13** (A) Mean of all the 1431 SMASSII asteroid spectra. (B) Principal components (PCs) derived from 1431 SMASSII asteroid spectra data. Their contribution ratios are ~0.79, 0.20, and 0.01 for PC1, PC2, and PC3, respectively. Note that nearly the same PCs can be given when the same data are decomposed after standardization and then multiplied with the standard deviations. (C) S-type bright boulders on Ryugu, anhydrous bright boulders on Bennu, laser-irradiated experiments, and different grain size data of ordinary chondrites in the PC1 and PC2 space. Crosses are asteroids from SMASSII data (Bus and Binzel, 2002; Binzel et al., 2004). Light blue circles indicate S-type bright boulders on Ryugu, lime circles indicate anhydrous bright boulders on Bennu, and site numbers designated by DellaGiustina et al. (2021) are also shown in the figure. Dashed lines are ordinary chondrites with pulsed laser irradiation (RELAB, Brown University).

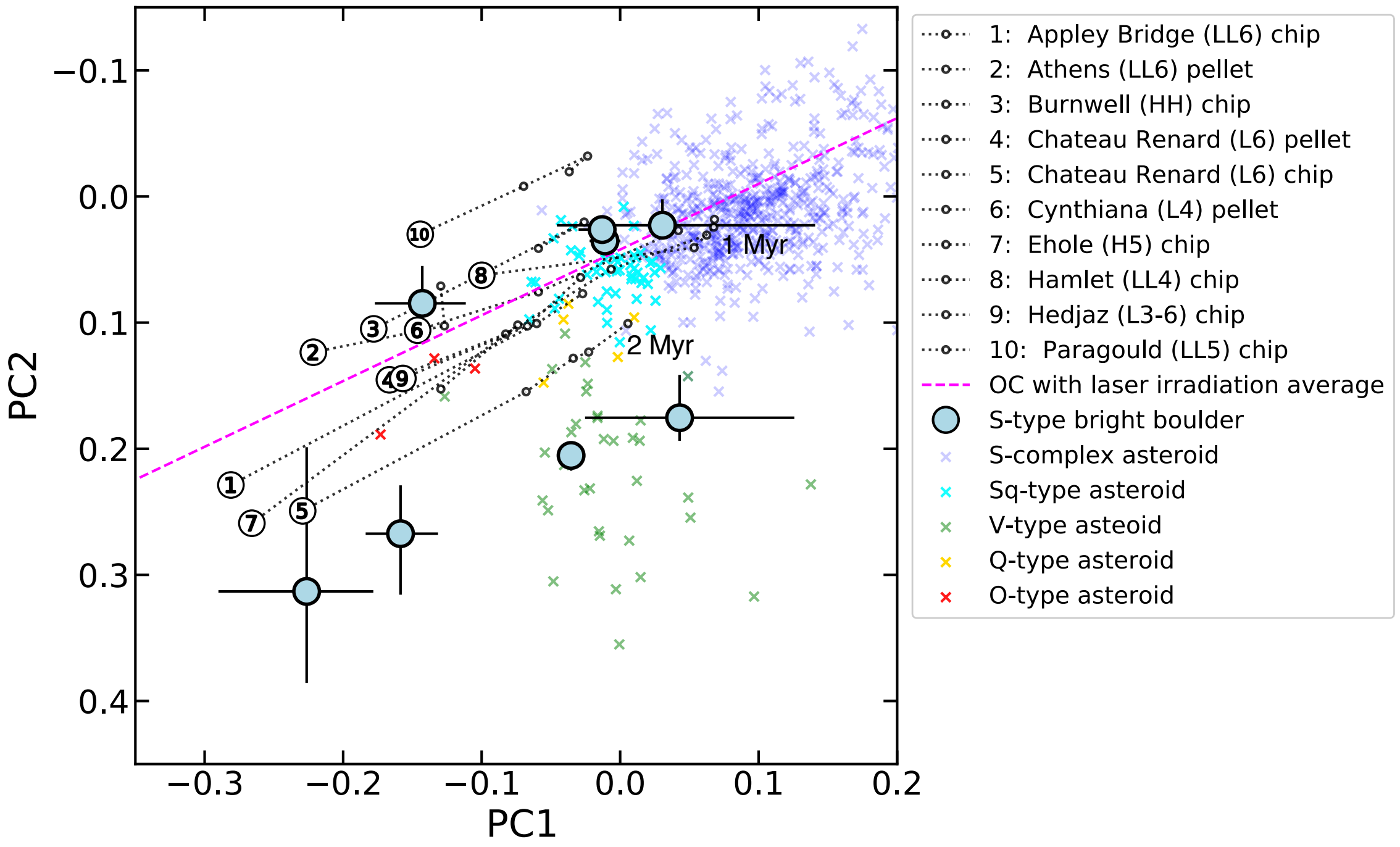

**Fig. 14** Laser-irradiated ordinary chondrites on PC1 and PC2 space (dashed lines). The largest white circles indicate unirradiated ordinary chondrites and the following plots are of the same meteorite sample with increasing energies of pulsed-laser irradiation (RELAB, Brown University; details of these meteorite samples are provided in Appendix A). The ID numbers correspond to meteorites in Table A.1. Estimated irradiation times of 1 Myr and 2 Myr correspond to Chateau Ranard (pellet, #4) with laser energy of 35 mJ and Chateau Ranard (Chip, #5 with laser energy of 80 mJ, respectively.

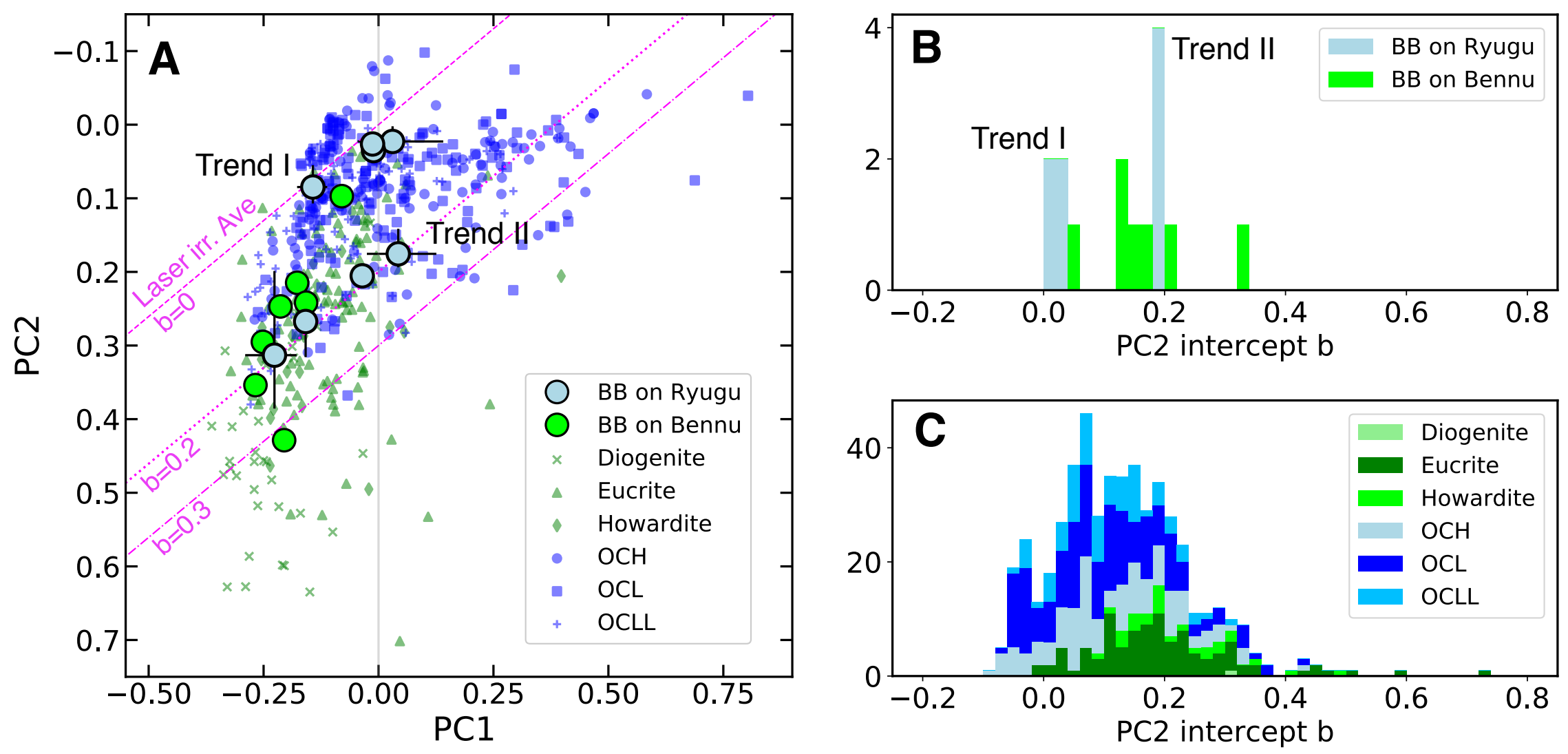

**Fig. 15** (A) Bright boulders on Ryugu and Bennu as well as meteorites in the PC1 and PC2 space. Trend-I and Trend-II bright boulders on Ryugu (light blue circles), bright boulders on Bennu (lime circles), ordinary chondrites (blue circles: H, blue squares: L, blues crosses: LL), and HED meteorites (green crosses: Diogenite, green triangles: Eucrite, green diamonds: Howardite) are shown. Magenta lines indicate the line with the slope in PC1-PC2 space ($a$=-0.52), which is the average of 10 laser-irradiated chondrites in Fig. 14. Dashed, dotted, and dash dotted line indicate the line with PC2 intercept $b$=0, 0.2, and 0.3, respectively. (B) (C) PC2 intercepts of bright boulders, ordinary chondrites, and HED meteorites in Fig. 15A. PC2 intercepts were calculated using the average slope $\alpha \sim -0.52$ in PC1 and PC2 space of laser-irradiated meteorites in Fig. 13A.

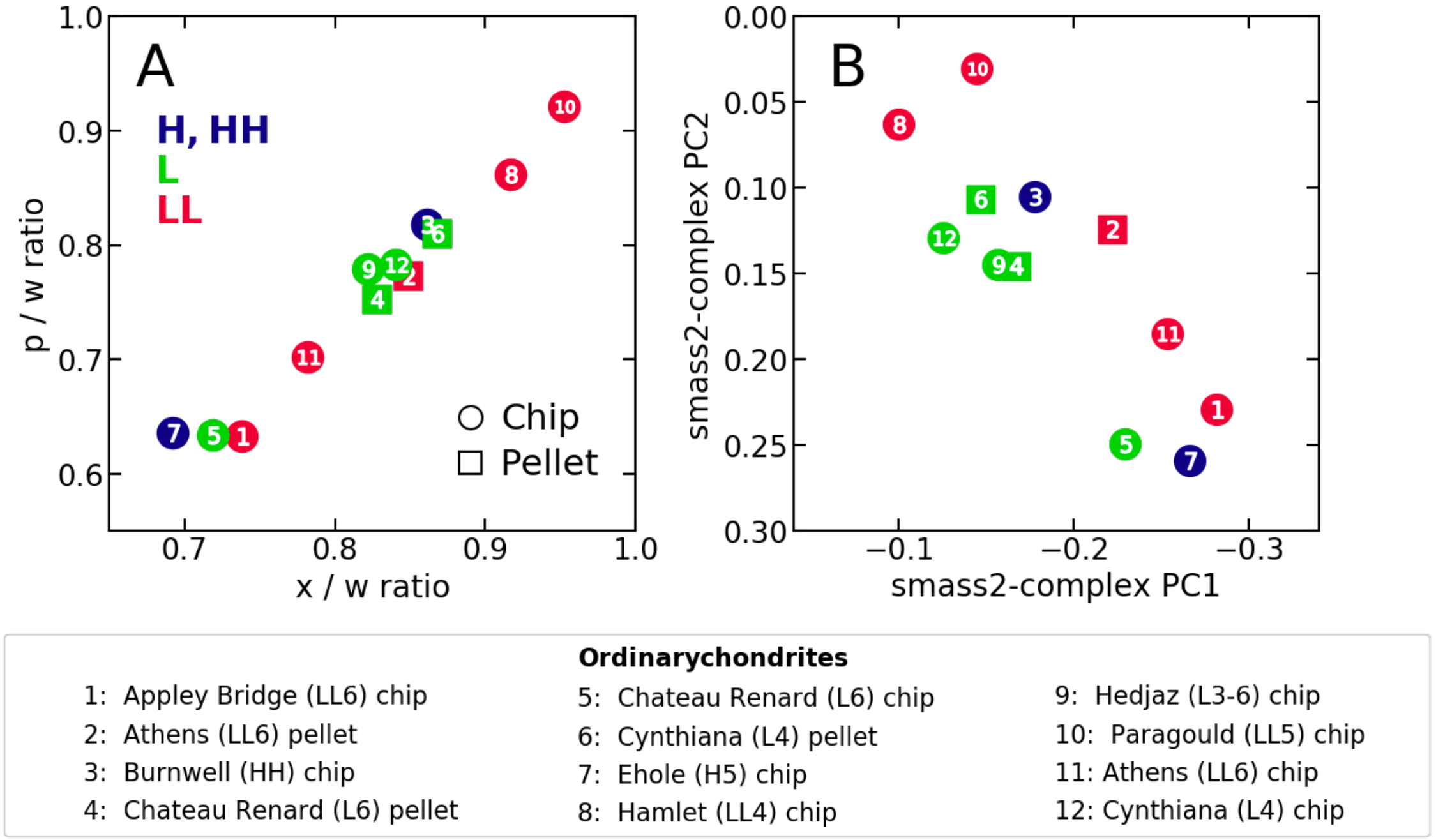

**Fig. 16** Longward absorptions and PC scores of unirradiated ordinary chondrites. (A) W, x, and p are the reflectance at 700, 860, and 950 nm. The ID numbers correspond to meteorites in Table A.1. The colors on the plots indicate the types of ordinary chondrites (H and HH: blue, L: lime and LL: red) and the shapes indicate the forms of samples (Circle: chip and square: Pellet) (B) Decomposed spectra of ordinary chondrites on PC1 and PC2 space. Colors and shapes of plots indicates types and sample forms as same as in (A).

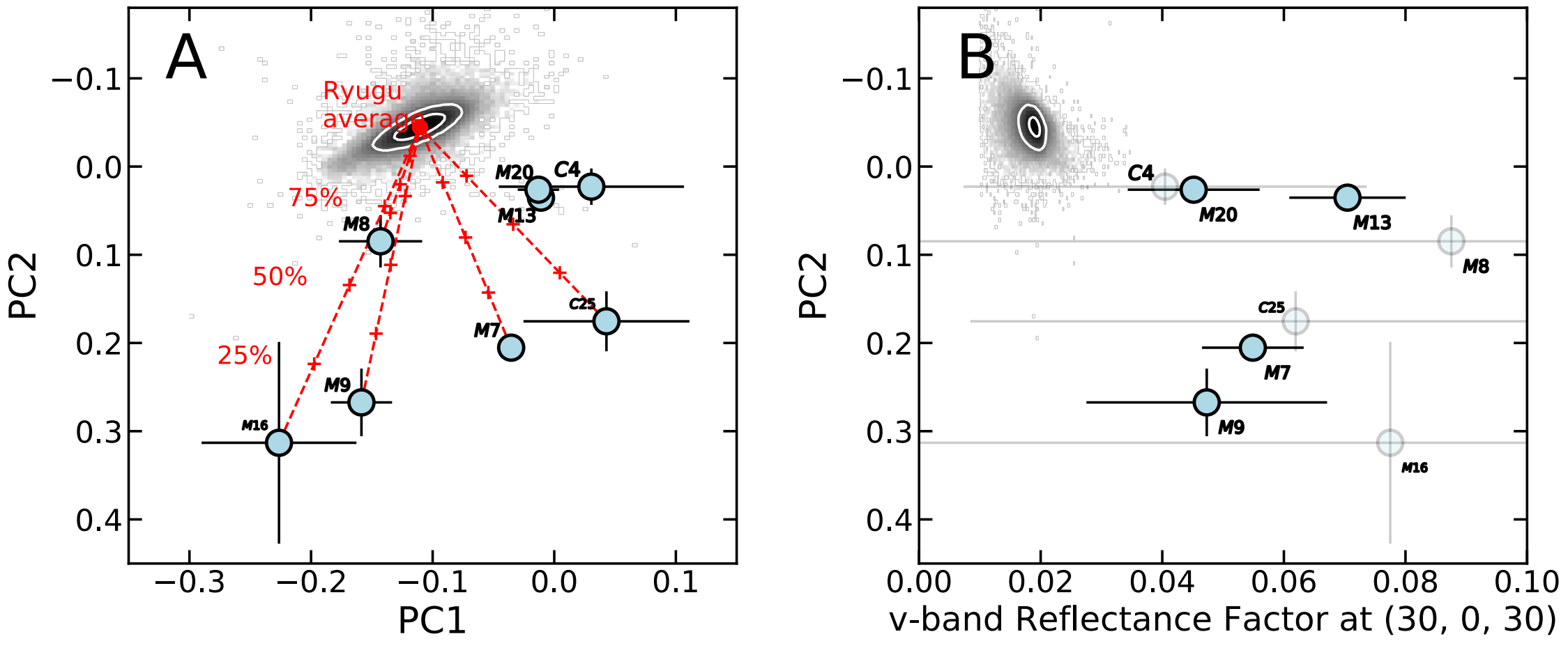

**Fig. 17** (A) Linear combinations of the spectra of S-type bright boulders and the global average of Ryugu (Sugita et al. 2019) in the PC1 and PC2 space (Red dotted lines). White contours indicate the surface of Ryugu, representing 95% and 68% of the surface area. (B) V-band reflectance factor and PC2 scores of S-type bright boulders at incidence, emission, and phase angles are 30°, 0°, and 30°, respectively.

**4.2 Spectral variety among C-type bright boulders**

Tatsumi et al. (2021) suggested an endogenic origin for the C-type bright boulders on Ryugu. First, C-type bright boulders are much more abundant (i.e., by a factor of ~10 for boulder sizes < 0.3m) than S-type bright boulders (Fig. 2 in Paper 1). Second, the largest bright boulder, M6 in Tatsumi et al. (2021), is one of the type-3 (bright and mottled) boulders found by Sugita et al. (2019). Most of the other C-type bright boulders in Tatsumi et al. (2021) exhibit spectra similar to these type 3-boulders, whose spectral trend is continuously connected to the global spectrum of Ryugu. These observations suggest that these C-type bright boulders are from the same parent body as Ryugu.

The spectral variation among C-type bright boulders can help us understand their origin. Tatsumi et al. (2021) found that although most C-type bright boulders exhibit spectra relatively close to the global average of Ryugu, some C-type bright boulders exhibit spectra with a much stronger UV absorption and a larger spectral slope. If all C-type bright boulders are fragments of the parent body of Ryugu, their spectral variety should be explained by some spectral modification mechanisms of the original materials constituting Ryugu. One of the major candidates for Ryugu meteorite analogs is thermally metamorphosed carbonaceous chondrites (Sugita et al., 2019; Kitazato et al., 2019). Results of laboratory experiments suggest that the colors of these chondrites are changed by various mechanisms, such as dehydration, space weathering, and grain size effects (Hiroi et al., 1996a; Hiroi et al., 1996b; Lantz et al., 2017; Cloutis et al., 2018). Tatsumi et al. (2021) compared C-type bright boulders and heated carbonaceous chondrites, finding that such color variation among C-type bright boulders could be explained by different degrees of thermal metamorphism. However, owing to their great spectral variety, whether it is a unique explanation based on a dozen of samples is difficult to conclude. Furthermore, no other likely processes, such as space weathering or grain size effects, have been examined in sufficient detail.

To further understand the spectral variety of C-type bright boulders and examine which color modification process can best account for their variety, we compared the spectral variety of bright boulders and laboratory experimental data of those effects on meteorites, using the spectra of the much larger number of C-type bright boulders from this study.

As the analysis method is similar to that used by Tatsumi et al. (2021), we only provide a short description of the method here. First, we discretized the continuous spectra of meteorites in the RELAB dataset into the ONC-T wavelengths. Following the analysis methods of Bus and Binzel (2002), the 4-band (ul: 0.40 $\mu$m, b: 0.48 $\mu$m, v: 0.55 $\mu$m, x: 0.86 $\mu$m) spectra are then fitted with the equation below using the least squares method.

$$f_i = 1 + \gamma(\lambda_i - 0.55), \tag{4.2.1}$$

where $f_i$ is the normalized reflectance at each band, $\lambda_i$ is the wavelength of the band in $\mu$m, and $\gamma$ is the slope of the fitted line, constrained to unity value at 0.55 $\mu$m. Second, we evaluate the UV index, which is a measure of the depth of the UV absorption. We evaluate the deviation in the UV from an extrapolated point based on the continuum spectral slope. The depth of the UV absorption $D_{ul}$ is defined as:

$$D_{ul} = \frac{ul}{ul_0} - 1, \tag{4.2.2}$$

where $ul_0$ is the extrapolated reflectance at the ul band wavelength from the v-to-x slope and $ul$ is the observed reflectance in the ul band. Thus, a positive $D_{ul}$ indicates a UV upturn, while a negative $D_{ul}$ indicates UV absorption. We evaluated the 4-band spectra of 15 C-type bright boulders observed during the 2.7-km hovering observation and 77 C-type bright boulders observed during the 1.7-km scanning observations. Note that spectrally ambiguous bright boulders C1 and C6 are also analyzed as C-type bright boulders in this section.

In the following subsections, we first describe the spectral variety among all the C-type bright boulders and then compare them with the spectra of meteorites that have experienced various processes that have been demonstrated to modify color properties.

The results of the comparison reveal that the continuous spectral trend found among C-type bright boulders exhibits the greatest similarity to the heating trends of Murchison (CM2) meteorites and is not consistent with Allende (C32) meteorites. We also found that other processes, such as space weathering and the grain size effect, do not account for their spectral variation as a sole source of the variations. Thus, their contribution may be secondary. These results suggest that thermal metamorphism may be the dominant cause for the spectral variety among C-type bright boulders. In addition, the change in albedo and spectral slope of some carbonaceous chondrites with low albedos, such as CM and CI chondrites, suggests that different degrees of dehydration could account for both the color and albedo variation among C-type bright boulders on Ryugu.

*Color variation of C-type bright boulders*: The slope/UV-absorption statistics of C-type bright boulders are shown in Fig. 18. Although analysis results by Tatsumi et al. (2021) potentially identify two groups in spectral slope (a group with spectra redder than the Ryugu's average and a group with bluer spectra), the greater number of bright boulder spectra analyzed in this study shows that the actual distribution is rather continuous. Several dozens of bright boulders newly analyzed in this study exhibit spectral slopes in the intermediate range between the two possible groups suggested by Tatsumi et al. (2021), indicating a continuous spectral trend of C-type bright boulders. It is also notable that this continuous trend overlaps with the global average spectrum of Ryugu's surface, which is consistent with the hypothesis that they are endogenous.

The entire spectral trend of all C-type bright boulders exhibits a V-shape in Fig. 18. The spectral slope of the Ryugu's globally averaged spectrum is ~0.1 $\mu m^{-1}$. Bright boulders with a spectral slope redder than the Ryugu's globally averaged spectrum exhibit a wider range in UV absorption than the bluer bright boulders. The range of the former is approximately twice that of the latter.

*Comparison with heated carbonaceous chondrites*: Laboratory experiments have shown a decrease in UV absorption upon heating for both CM and CI chondrites, making their spectral slope flatter (Hiroi et al., 1993; Hiroi et al., 1996a, 1996b). To assess this possibility, we compared C-type bright boulder spectra with spectra from heating experiments on carbonaceous chondrites. We selected major carbonaceous chondrites widely studied in the literature: Murchison (CM2), Ivuna (CI1), and Allende (CV3).

Among the three chondrites, Murchison spectra exhibit changes in the slope/UV-index properties most similar to the distribution of C-type bright boulders (Fig. 18). In the heating experiments of Murchison, the UV index first increases until heated to temperatures up to ~700°C and then decreases when heated to higher temperatures. The fist increase in UV absorption at lower temperature are ascribed to the decomposition of hydrated minerals and carbonization of the organic material and subsequent decrease at higher tempretures may be due to the formation of Fe-bearing secondary olivine and small FeNi metal particles and carbon evaporation (Nakamura et al., 2005; Applin et al., 2018). Furthermore, this spectral change in carbonaceous chondrite due to heating is generally continuous. Thus, the heating track of Murchison exhibits a V-shaped movement in this UV-index/v-to-x slope diagram, which coincides with the continuous spectral trend of the C-type bright boulders on Ryugu. Note that the spectra of surfaces on Ryugu also are distributed along the heating track of Murchison at ~ 700–900 °C in this diagram, which is consistent with PCA results by Sugita et al. (2019). This suggests that the color variations among C-type bright boulders and the average surface color of Ryugu might reflect the different degrees of dehydration of the same material and that such material might follow a similar color modification process to that of Murchison.

In contrast, the spectra of the two other chondrites show no apparent similarity in the changes in the UV-index/v-to-x slope space correlating with the C-type bright boulders (Fig. 18). The UV index of Ivuna increases with bluing in the v-to-x slope while being heated up. Due to its straight shape, this heating track is not very consistent with the color variation seen in bright boulders on Ryugu. However, it is noted that the Ivuna experiment ends at 700 °C; movement in UV-index/v-to-x slope space that would occur at higher temperatures is not currently known. If the spectral change of CI meteorites exhibits a different trend from the lower temperature range, as observed in CM meteorites, then the overall spectral change during the heating process may become more consistent with C-type bright boulders on Ryugu. Thus, the possibility for CI meteorites should not be excluded. Furthermore, comparison in the slope-reflectance space indicates that heated CI samples resemble the bright boulder distribution more than the heated CM samples (Fig. 19). Nevertheless, the great similarity to the CM heating track is difficult to match with the CI heating track, even if the higher temperature range has a more compatible spectral trend.

In contrast, the spectral changes along the heating track for Allende is much different from Ryugu's C-type bright boulders; it shows little spectral change over a wide range of temperatures (from unheated to ~ 900 °C). It also shows a negative UV index, which is not consistent with the surface of Ryugu or with the UV upturn seen in the bright boulder spectra.

Heating also changes the albedo of carbonaceous chondrites. In Fig. 19, we show a comparison of 0.55-$\mu$m reflectance factor and spectral slope of bright boulders with those of heated carbonaceous chondrites. Note that only bright boulders with areas ≧ 10 pixel$^2$ are shown here because the reflectance values of these well-resolved bright boulders are determined with high precision. General boulders with an albedo close to the Ryugu's average studied by Sugita et al. (2019) are also shown in this figure.

Among the three carbonaceous chondrites, the heating tracks of Ivuna and Murchison are relatively consistent with bright boulders and the general boulders on Ryugu. General boulders on Ryugu show a continuous trend from bright blue to dark red (Fig. 19). Their distribution trend is consistent with the heating track of Ivuna from 100–700 °C. The reflectance of heated Murchison is consistent with bright boulders with a redder spectral slope but not consistent with bright boulders with a bluer spectral slope or the general boulders. Allende is not consistent with bright boulders nor general boulders because of their high reflectance. Although neither Murchison nor Ivuna provide an exact match with bright boulders, their similarity to the bright boulders is much greater than that for Allende meteorites.

Furthermore, it is noted that the albedo comparison between Ryugu bright boulders and heated carbonaceous chondrites suggests that both bluer and redder C-type bright boulders could be from the same carbonaceous chondrite. When Murchison and Ivuna are heated, their albedo and spectral slope decrease at lower temperatures and then start to increase at high temperatures (Hiroi et al., 1996a, 1996b). This reversal in albedo and spectral slope can be observed in our albedo/spectral slope diagram as well (Fig. 19). If the average reflectance of Ryugu is close to the turnaround point (i.e., darkest point) of the spectral heating track, as suggested by Sugita et al. (2019), the spectra of samples heated at both higher and lower temperatures would a have higher reflectance than the average of the asteroid. Furthermore, both redder and bluer bright boulders could be explained by different degrees of dehydration. In other words, different degrees of dehydration of CM and CI meteorites could account for both the color variation of C-type bright boulders and their higher reflectance. Another important finding from the spectral comparison is that both heated CM and CI chondrites, whose spectral properties overlap with the bright boulders on Ryugu, exhibit the closest spectra to the average Ryugu spectrum at medium (i.e., several hundred °C) temperatures. This observation supports the notion that the bulk of Ryugu's materials may have experienced moderate thermal metamorphism.

*Comparison with laser-irradiated carbonaceous chondrites and grain size effect*: Recent laboratory experiments on space weathering of carbonaceous chondrites indicate that their spectral modification trend may be much more complicated than for ordinary chondrites (e.g., Lantz et al., 2017; Nakamura et al., 2020). Several experiments suggest that pulsed laser irradiation on Murchison results in an increase in the UV index, as shown in Fig. 20. In contrast, irradiation on a dehydrated chondrite (Dhofar 225) leads to reddening of the spectral slope. Smaller grain sizes also lead to redder spectra (Cloutis et al., 2018). For example, the range in the v-to-x slope in Murchison samples of different grain size is about half that of the C-type bright boulders (Fig. 20). As these effects induce changes in chondrite spectra in only one spectral direction, none of these single processes can explain the entire spectral variation among Ryugu's C-type bright boulders. Furthermore, the V-shape distribution of bright boulders in the slope/UV-index space is difficult to explain with the combination of these two effects (i.e., reddening and decrease in UV-absorption) because these spectral modification mechanisms are not known to produce the observed change in spectral trend as they proceed. However, they could make a significant contribution to the widening of the range of the spectral distribution of bright boulders. Thus, these effects may have to be considered for a detailed spectral analysis of C-type bright boulders on Ryugu.

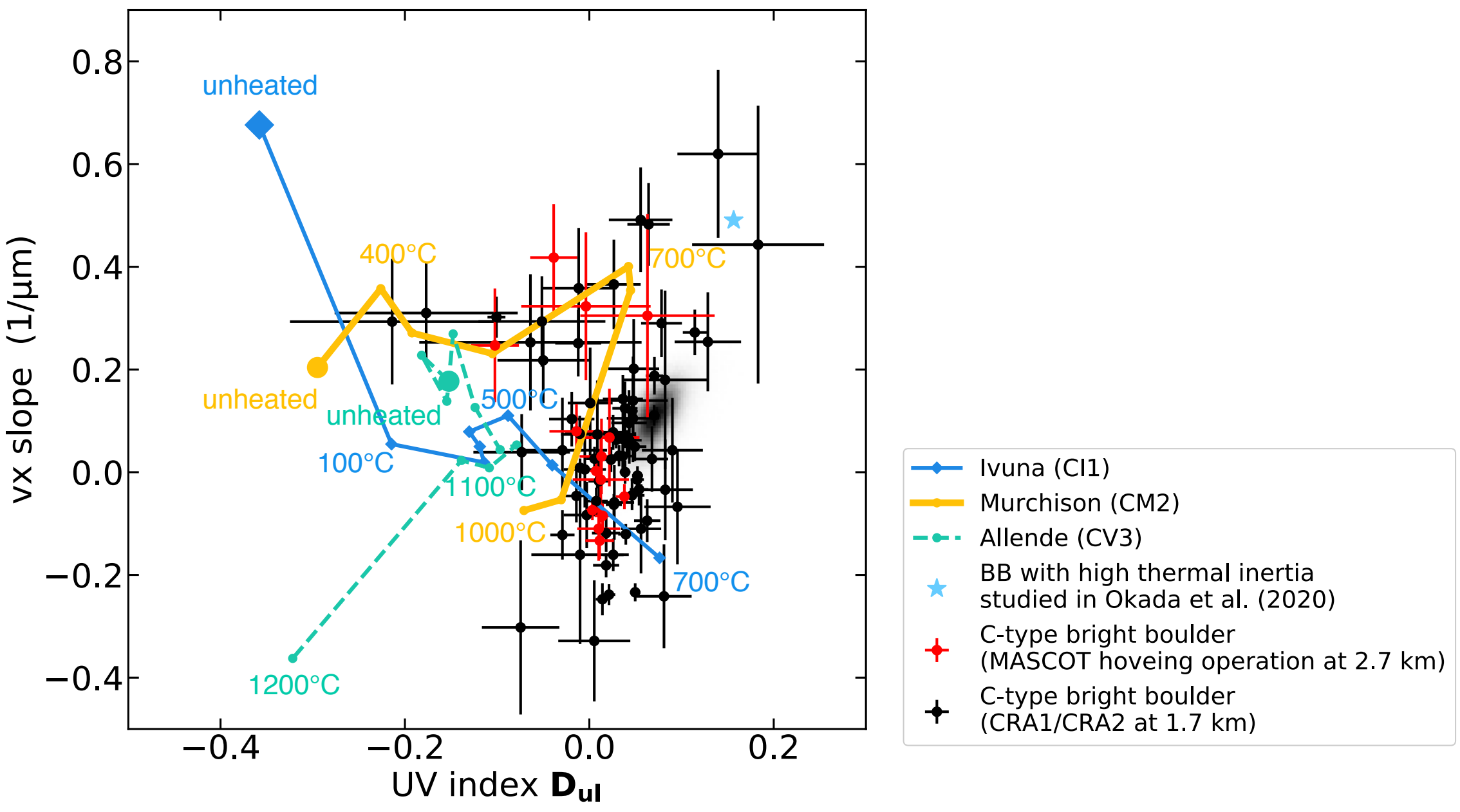

**Fig. 18** Comparison of UV-index and spectral slope between C-type bright boulders on Ryugu and heated carbonaceous chondrites. Red circles are bright boulders observed during the 2.7-km hovering observation and black circles are bright boulders observed during the 1.7-km scanning observations. Colored lines indicate the heating tracks of carbonaceous chondrites. Large symbols indicate unheated samples and small symbols indicate samples heated at various temperatures. Blue diamonds indicate Ivuna (unheated and heated at 100–700 °C), yellow circles indicate Murchison (unheated and heated at 400–1000 °C), and green circles with a dashed line indicate Allende (unheated and heated at 400–1200 °C) taken from the RELAB database. The black cloud indicates the spectral distribution for the surface of Ryugu (Sugita et al. 2019). Light blue star indicates the bright boulder with high thermal inertia, which is studied in Okada et al. (2020).

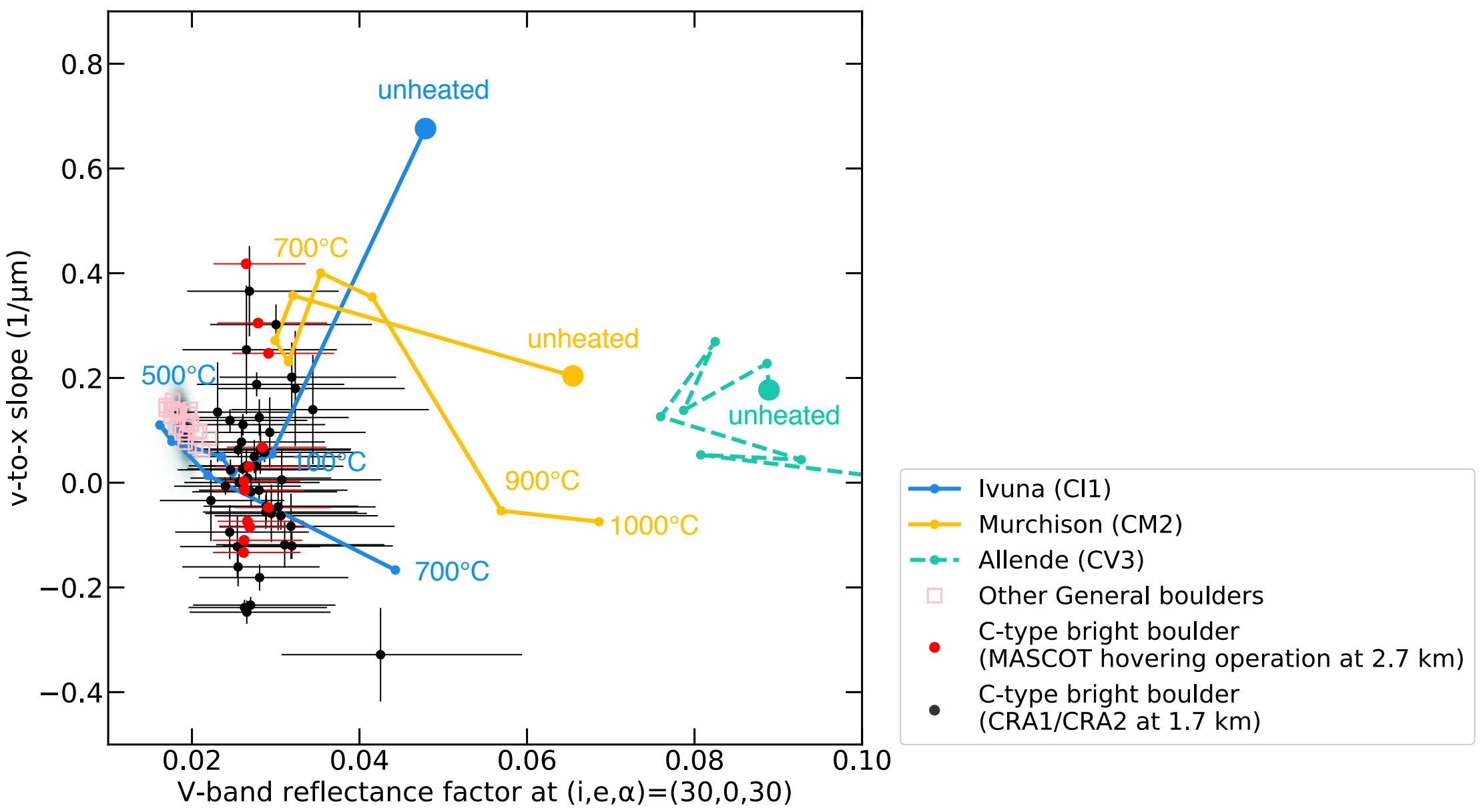

**Fig. 19** Color comparison of spectral slope and 0.55-$\mu$m reflectance factor at incidence, emission, phase angles (i, e, α) = (30°, 0°, 30°) between C-type bright boulders, other general boulders on Ryugu, and heated carbonaceous chondrites. Pink open squares indicate 22 other large general boulders (Type1 and Type2 boulders) studied in Sugita et al. (2019). Colored symbols connected with solid or dashed lines indicate the same heating tracks of meteorites as in Fig 18. Red circles are bright boulders observed during the 2.7-km hovering observation and black circles are bright boulders observed during the 1.7-km scanning observations. Note that only bright boulders with surface area ≧10 pixel$^2$ are shown here because the reflectance of bright boulders with a surface area <10 pixel$^2$ cannot be evaluated with sufficiently high precision because of their small sizes.

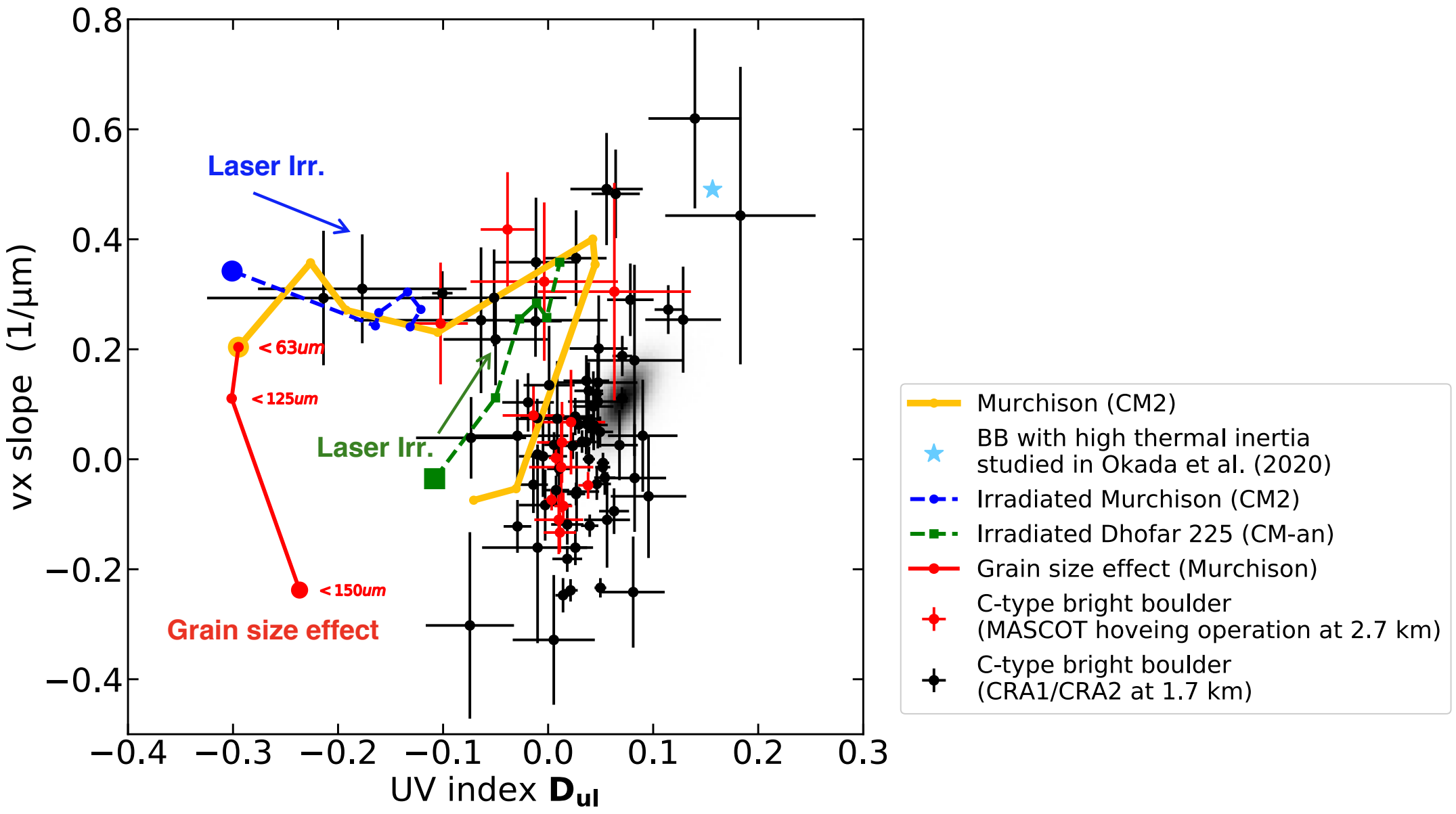

**Fig. 20** Comparison of UV-index and spectral slope of between C-type bright boulders, carbonaceous chondrites with laser irradiation, and carbonaceous chondrites of different grain sizes. Blue circles and green squares with dashed line are irradiation tracks of Murchison and Dhofar 225, respectively. Red circles with solid line are the samples of Murchison with different grain sizes. Red circles are bright boulders observed during the 2.7-km hovering observation, and black circles are bright boulders observed during the 1.7-km scanning observations. The yellow line for the heating track of Murchison and black cloud for the Ryugu surface distribution are the same as in Fig. 18.

### 4.3 Implications for the evolution of Ryugu from its parent body

In this section, we discuss the geologic implications of the newly found spectral and morphologic properties of bright boulders on Ryugu. More specifically, we discuss the thermal history of Ryugu's parent body in Section 4.3.1 and the collisional break-up history of Ryugu's parent body in Section 4.3.2.

*4.3.1 Thermal history of the parent body*: C-type bright boulders are too large to have accumulated after Ryugu formed into its current state (Tatsumi et al., 2021). The spatial distribution and spectral variation among C-type bright boulders on Ryugu are thus useful for understanding the thermal history of these fragments. As discussed in Section 4.2, the color variation among C-type bright boulders is more likely to reflect the different degrees of thermal metamorphism experienced by these boulders than differences in their maturity due to space weathering and grain size variations. In addition, their spatial distribution could constrain the timing of such dehydration. The ubiquity of bright boulders on Ryugu suggests that bright boulders were mixed into Ryugu a long time ago, during the rubble-pile accumulation, although more recent mixing cannot be ruled out if granular convection on Ryugu is highly efficient (Paper1). Thus, the dehydration process which induced the color variations in C-type bright boulders is likely to have occurred within Ryugu's parent body. In other words, the different spectral properties among the C-type bright boulders may reflect different components within the parent body, which have experienced dehydration to different degrees than the average Ryugu materials.

Furthermore, the size distribution of bright boulders allows us to infer the mass ratio of the various components that experienced different degrees of thermal metamorphism in Ryugu's parent body. Similar power−law indexes at different latitude ranges on Ryugu suggest that the size frequency of bright boulders is globally uniform and that the area ratios of bright boulders to general boulders is uniformly low (~0.1% as discussed in Paper 1) on the surface of Ryugu. This indicates that the majority of the boulder population on Ryugu displays homogeneous spectra (general boulders), and only a small portion of the boulder population comprising Ryugu has distinctive spectra (bright boulders). If spectral difference between the average Ryugu material and bright boulders results from the degrees of thermal metamorphism as suggested in Section 4.2, then the observed spectral uniformity in the large mass of Ryugu's materials would suggests that Ryugu's parent body experienced homogenous thermal metamorphism (i.e., general boulders) with minor exceptions of bright boulder materials. This highly homogeneous thermal metamorphism and small mass fraction of material processed to low degrees of thermal metamorphism are consistent with internal heating in a large parent body due to radiogenic heating (i.e., scenario 1 by Sugita et al., 2019).

Here, it is important to note that the large-scale impact expected for catastrophic disruption of the parent body could generate a large volume of thermal metamorphism. Recent numerical calculations show that an impact at an average collision speed (~5 km/s) among MBAs can lead to several hundred degrees of impact heating on local zones of the parent body assuming it was highly porous (Ballouz et al., 2020; Michel et al., 2020; Jutzi and Michel, 2020). This level of impact heating may lead to moderate degrees of dehydration as suggested by Kitazato et al. (2020) and Sugita et al. (2019). Furthermore, because such an impact would leave a large fraction of parent-body materials largely unheated, it can generate rubble piles similar to Bennu containing a high abundance of hydrated minerals (Lauretta et al., 2019; Hamilton, 2019). If such shock-induced dehydration is the main mechanism to generate bright boulders from average Ryugu materials, shock comminution and fragmentation would increase porosity. Thus, the thermal inertia of bright boulders would be expected to be lower than average. However, high-resolution observations with the thermal infrared camera (TIR) indicate that a bright boulder (Fig. 5a in Paper 1), whose spectral properties are more consistent with lower thermal metamorphism (Fig. 18), has a high thermal inertia (i.e., low porosity) (Okada et al., 2020). Also, boulder materials on Bennu have similar thermal inertia as on Ryugu (DellaGiustina and Emery et al., 2019). Furthermore, an impact would lead to highly heterogenous heating (e.g., Davison et al. 2013; Sugita et al. 2019). As reaccumulated impact fragments tend to collect materials originally located at a wide range of positions along the streamlines of excavation flow (Michel et al., 2015; Sugita et al., 2019; Michel et al., 2020; Ballouz et al., 2020), they could experience a wide range of impact heating. This is rather inconsistent with the observed homogeneity of Ryugu's materials. Nevertheless, the mass of Ryugu is only $10^{-6}$ of the probable parent body (~100 km in diameter). The calculations by Michel et al. (2020) and Ballouz et al. (2020) also show the scale of heterogeneity in shock heating for a catastrophic disruption impact of a 100-km parent body is on the order of 10 km, which is much larger than the size of Ryugu (~1 km). Thus, resulting re-accumulated rubble piles may collect fragments that experienced similar degrees of impact heating. The key issue is whether mixing along the excavating streamline is efficient enough to have a small rubble pile to represent the large parent body. The resolutions of numerical calculations on the catastrophic disruption thus far is not high enough yet to reach a decisive conclusion on this process. Thus, higher-resolution studies and detailed geochemical analyses of the Ryugu samples to be returned to Earth are needed.

*4.3.2. Collisional history*: The results from our spectral analysis of S-type bright boulders also have important implications for Ryugu's collisional history. Through comparison of the spectral properties between S-type bright boulders on Ryugu, bright boulders on Bennu, and meteorites, the spectral variety of all S-type bright boulders is found to be consistent with two separate projectiles, and Ryugu and Bennu have experienced different collisional histories and perhaps had different parent bodies. More specifically, Ryugu may be from the catastrophic disruption of the parent body of Polana or Eulalia and the subsequent reaccumulation. However, the breakup ages of Polana (i.e., ~1.4 Gyr) and Eulalia (i.e., ~0.8 Gyr) are significantly longer than the collisional lifetime of Ryugu based on classical scaling rules (i.e., ~several 100 Myr) (Bottke et al., 2015). Thus, Ryugu may not have been formed directly from Polana's parent body or Eulalia's, but it may have been formed from the catastrophic disruption of a member of the Polana or Eulalia family (Sugita et al., 2019; Walsh et al., 2020).

Furthermore, as discussed in detail by Paper 1, the fact that both C-type and S-type bright boulders are mixed within breccias as intra-boulder clasts suggests that these clasts were generated (i.e., fragmented) or delivered to Ryugu's parent body when cementation processes were still active. Because exogeneous bright boulders are found as clasts within breccias on Bennu, a similar process must have occurred on Bennu's parent body. Thus, if the age of cementation is estimated from geochemical analyses of returned samples, the timings of breakups of Ryugu's and Bennu's parent bodies and delivery of projectile materials can be constrained.

## 5. Conclusions

In this study, we conducted spectral analysis of 12 sets of multi-color images using data obtained during the 1.7-km scanning observations. Using multiband, high-resolution (~0.18 m/pixel) images obtained by the ONC-T, we conducted the spectroscopic analyses of an additional 79 bright boulders. Only 3 of these bright boulders overlap with the 21 bright boulders studied by Tatsumi et al. (2021). The analysis of bright boulders, as well as more than a thousand bright boulders (brightness ≧ 1.5 times the surrounding regolith and boulders) detected in our measurements, yielded the following conclusions:

- S-type bright boulders on Ryugu follow two distinctively different parallel trends in PC space ranging from ordinary chondrite spectra, which would represent extremely fresh asteroid surfaces that has not experienced significant space weathering, to S-, Q- and V-type MBA spectra.
- The gap between the two parallel trends among S-type bright boulders on Ryugu cannot be readily accounted for by the effects of grain size or dust covering, but it is consistent with the spectral variations among different compositions of ordinary chondrites.
- These two trends are found to be parallel to the space-weathering tracks simulated through laser irradiation experiments on ordinary chondrites. This result strongly suggests that the variation observed along the color trend among S-type bright boulders reflects different maturity levels due to space weathering.
- The laser irradiation dose required to mature ordinary chondrites to the observed spectra of S-type boulders on Ryugu is equivalent to $10^5$–$10^6$ years of exposure to space at Ryugu's current orbit. This finding is consistent with the retention age (<$10^6$ years) of ~10-m-diameter craters, which would represent the age of the top ~1-m surface layer on Ryugu (Sugita et al., 2019; Morota et al., 2019) based on a gravity-scaling rule for granular materials by Tatsumi and Sugita (2018). This agreement strongly supports the interpretation that Ryugu's surface is extremely young, suggesting that the samples acquired from Ryugu's surface will be fresh.
- An examination of the large S-type clast embedded in large breccia indicated no serpentine absorption at 0.7 $\mu$m. This implies that the fragmentation and cementation of this breccia occurred after the termination of aqueous alteration, as proposed in Paper 1.
- Spectral analysis of 90 C-type bright boulders reveals that they have a continuous spectral trend similar to the heating track of low-albedo carbonaceous chondrites, such as CM and CI. This indicates that the two trends among C-type bright boulders by Tatsumi et al. (2021) are actually parts of one continuous V-shaped trend.

- The spectra of other heated carbonaceous chondrite samples, such as Allende (CV3), do not match the observed spectral trend among C-type bright boulders on Ryugu.
- Other processes, such as space weathering and grain size effect, cannot account for the spectral variation in C-type bright boulders on their own.

These results have important implications for the evolution of Ryugu's parent body:

- Assuming that a single S-type asteroid does not have two distinctively different compositions of materials, Ryugu's parent body would have been hit by more than one large projectiles during its history.
- The similarity in the spectral trend of C-type bright boulders to the heating track of CM chondrites supports the evolution scenario, that is, the parent body experienced thermal metamorphism. The low abundance ratio (~$10^{-5}$ in volume) of C-type bright boulders presented shown in Paper 1 suggests that the majority of boulders on Ryugu are characterized by homogeneous spectra and only a small portion of Ryugu's boulders display distinctive spectra. These pieces of evidence support the internal heating rather than impact heating.

**Acknowledgments.** The authors wish to thank all members of JAXA's Hayabusa2 team. The reflectance spectra of meteorite samples used in this study are from the RELAB database. This study was supported by the Japan Society for the Promotion of Science (JSPS) Core-to-Core Program "International Network of Planetary Sciences."

D. Domingue was supported by the NASA Hayabusa2 Participating Scientist Program (NNX16AL34G), and the Solar System Exploration Research Virtual Institute 2016 (SSERVI16) Cooperative Agreement (NNH16ZDA001N) SSERVI-TREX. P.M. acknowledges funding from the French space agency CNES, from Academies of Excellence: Complex systems and Space, environment, risk, and resilience, part of the IDEX JEDI of the Université Côte d'Azur, and from the European Union's Horizon 2020 research and innovation programme under grant agreement No 870377 (project NEO-MAPP).

# Appendix

**Table A.1** Ordinary chondrite samples from RELAB archive

| Spectra ID | Sample Name | | Grain size $\mu$m | Laser irradiation mJ | ID number in Fig. 4.1.2 |
|---|---|---|---|---|---|
| C1OC12A | Appley Bridge (LL6) | chip | | | 1 |
| C1OC12A20 | Appley Bridge (LL6) | chip | | 20 | |
| C1OC12A40 | Appley Bridge (LL6) | chip | | 40 | |
| C1OC13D | Athens (LL6) <125 um | pellet | < 125 | | 2 |
| C1OC13D05 | Athens (LL6) <125 um | pellet | < 125 | 5 | |
| C1OC13D15 | Athens (LL6) <125 um | pellet | < 125 | 15 | |
| C1OC21A | Burnwell (HH) chip | chip | | | 3 |
| C1OC21A20 | Burnwell (HH) | chip | | 20 | |
| C1OC21A40 | Burnwell (HH) | chip | | 40 | |
| C1OC11D | Chateau Renard (L6) | pellet | < 125 | | 4 |
| C1OC11D05 | Chateau Renard (L6) | pellet | < 125 | 5 | |
| C1OC11D15 | Chateau Renard (L6) | pellet | < 125 | 15 | |
| C1OC11D35 | Chateau Renard (L6) | pellet | < 125 | 35 | |
| C1OC11A | Chateau Renard (L6) | chip | | | 5 |
| C1OC11A20 | Chateau Renard (L6) | chip | | 20 | |
| C1OC11A40 | Chateau Renard (L6) | chip | | 40 | |
| C1OC11A60 | Chateau Renard (L6) | chip | | 60 | |
| C1OC11A80 | Chateau Renard (L6) | chip | | 80 | |
| C1OC15D | Cynthiana (L4) | pellet | < 125 | | 6 |
| C1OC15D05 | Cynthiana (L4) | pellet | < 125 | 5 | |
| C1OC15D15 | Cynthiana (L4) | pellet | < 125 | 15 | |
| C1OC06A | Ehole (H5) | chip | | | 7 |
| C1OC06A20 | Ehole (H5) | chip | | 20 | |
| C1OC06A40 | Ehole (H5) | chip | | 40 | |
| C1OC02A | Hamlet (LL4) | chip | | | 8 |
| C1OC02A20 | Hamlet (LL4) | chip | | 20 | |
| C1OC02A40 | Hamlet (LL4) | chip | | 40 | |
| C1OC02A60 | Hamlet (LL4) | chip | | 60 | |
| C1OC16A | Hedjaz (L3-6) | chip | | | 9 |
| C1OC16A20 | Hedjaz (L3-6) | chip | | 20 | |
| C1OC16A40 | Hedjaz (L3-6) | chip | | 40 | |
| C1OC07A | Paragould (LL5) | chip | | | 10 |
| C1OC07A20 | Paragould (LL5) | chip | | 20 | |
| C1OC07A40 | Paragould (LL5) | chip | | 40 | |
| C1OC07A60 | Paragould (LL5) | chip | | 60 | |
| C1OC13A | Athens (LL6) | chip | | | 11 |
| C1OC15A | Cynthiana (L4) | chip | | | 12 |